\documentclass{aastex63}

\usepackage{appendix}
\usepackage{hyperref}

\submitjournal{ApJ}

\shorttitle{\gigantes: Voids from the Quijote Simulations}
\shortauthors{Kreisch et al.}

\graphicspath{{./}{figures/}}

\newcommand{\AP}{Alcock-Paczy\'nski~}
\newcommand{\gigantes}{\texttt{GIGANTES}}
\newcommand{\quijote}{\texttt{QUIJOTE}}
\newcommand{\vide}{\texttt{VIDE}}

\begin{document}

\title{The \gigantes~dataset: precision cosmology from voids in the machine learning era}

\correspondingauthor{Christina Kreisch and Alice Pisani}
\email{ckreisch@astro.princeton.edu, apisani@astro.princeton.edu}

\author[0000-0002-5061-7805]{Christina D. Kreisch}
\affiliation{Department of Astrophysical Sciences, Princeton University, 4 Ivy Lane, Princeton, NJ 08544 USA}

\author[0000-0002-6146-4437]{Alice Pisani}
\affiliation{Department of Astrophysical Sciences, Princeton University, 4 Ivy Lane, Princeton, NJ 08544 USA}

\author[0000-0002-4816-0455]{Francisco Villaescusa-Navarro}
\affiliation{Department of Astrophysical Sciences, Princeton University, 4 Ivy Lane, Princeton, NJ 08544 USA}

\author[0000-0002-5151-0006]{David N. Spergel}
\affiliation{Department of Astrophysical Sciences, Princeton University, 4 Ivy Lane, Princeton, NJ 08544 USA}
\affiliation{Center for Computational Astrophysics, Flatiron Institute, 162 5th Avenue, New York, NY 10010 USA}

\author[0000-0002-5854-8269]{Benjamin D. Wandelt}
\affiliation{Center for Computational Astrophysics, Flatiron Institute, 162 5th Avenue, New York, NY 10010 USA}
\affiliation{Institut d'Astrophysique de Paris, UMR 7095, CNRS, 98 bis Boulevard Arago, 75014 Paris, France}
\affiliation{Sorbonne Universit\'es, Institut Lagrange de Paris, 98 bis Boulevard Arago, 75014 Paris, France}

\author[0000-0002-0876-2101]{Nico Hamaus}
\affiliation{Universit\"ats-Sternwarte M\"unchen, Fakult\"at f\"ur Physik, Ludwig-Maximilians Universit\"at, Scheinerstr.~1, 81679 M\"unchen, Germany}

\author[0000-0002-3568-3900]{Adrian E. Bayer}
\affiliation{Berkeley Center for Cosmological Physics, University of California, 341 Campbell Hall, Berkeley, CA 94720, USA}
\affiliation{Department of Physics, University of California, 366 LeConte Hall, Berkeley, CA 94720, USA}

\begin{abstract}

We present \gigantes, the most extensive and realistic void catalog suite ever released---containing over 1 billion cosmic voids covering a volume larger than the observable Universe, more than 20 TB of data, and created by running the void finder \texttt{VIDE} on \quijote's halo simulations. The expansive and detailed \gigantes~suite, spanning thousands of cosmological models, opens up the study of voids, answering compelling questions: Do voids carry unique cosmological information? How is this information correlated with galaxy information? Leveraging the large number of voids in the \gigantes~suite, our Fisher constraints demonstrate voids contain additional information, critically tightening constraints on cosmological parameters. We use traditional void summary statistics (void size function, void density profile) and the void auto-correlation function, which independently yields an error of $0.13\,\mathrm{eV}$ on $\sum\,m_{\nu}$ for a 1~$h^{-3}\mathrm{Gpc}^3$ simulation, without CMB priors. Combining halos and voids we forecast an error of $0.09\,\mathrm{eV}$ from the same volume. Extrapolating to next generation multi-Gpc$^3$ surveys such as DESI, Euclid, SPHEREx, and the Roman Space Telescope, we expect voids should yield an independent determination of neutrino mass. Crucially, \gigantes~is the first void catalog suite expressly built for intensive machine learning exploration. We illustrate this by training a neural network to perform likelihood-free inference on the void size function. Cosmology problems provide an impetus to develop novel deep learning techniques, leveraging the symmetries embedded throughout the universe from physical laws, interpreting models, and accurately predicting errors. With \gigantes, machine learning gains an impressive dataset, offering unique problems that will stimulate new techniques. 

\end{abstract}


\section{Introduction} \label{sec:intro}

Cosmic voids---the large under-dense regions infilling the cosmic web \citep{Gregory_1978, Joeveer_1978, Kirshner_1981,vandeweygaert_1993,peebles_LSS,Bond_1996b}---have gained interest in the last ten years as a robust tool to extract cosmological information \citep[see][and references therein]{Pisani_2019}. Until a few decades ago survey volumes were too small to provide enough statistics for voids, since voids are among the largest objects in the Universe. Large-scale surveys are now ushering in the era of big data in astronomy and astrophysics, which is essential for unveiling the power of cosmic voids. With such large amounts of data and access to previously unresolved phenomena, machine learning is becoming a prudent tool in astrophysics \citep[see e.g.][and references therein]{astro2020ml,mlinastro,imnn,anom,siyu,pacoml}.

Many traditional techniques, such as multi-layer perceptrons (MLPs) and convolutional neural networks (CNNs), are not interpretable, making discovering new physics a challenge, and can obscure information. The inherently graphical nature of the Universe has prompted substantial recent development in interpretable deep learning methods with graph neural networks (GNNs) \citep{Miles_2020}. {\color{black}At the same time, symmetries are embedded in physical laws that govern the evolution of the Universe. Cosmic voids offer more symmetries that can improve interpretable machine learning techniques.} The data-rich field of astrophysics, with its large surveys, offers unique problems fundamentally characterized by symmetries, interpretability, and the need for accurate error estimates, providing an impetus for the development of new machine learning techniques to conquer these questions.

With the advent of large modern surveys, we are entering the golden age of cosmic voids. 
The Baryon Oscillation Spectroscopic Survey (BOSS) \citep{BOSS_alam2017} and eBOSS \citep{eBOSS_dawson2016} have already provided several thousand voids \citep{Hamaus_2020,Aubert_2020}. This number will dramatically increase in the coming years thanks to data from the DESI experiment \citep{DESI2016}, the Subaru survey with the Prime Focus Spectrograph (PFS) \citep{pfs2016}, the SPHEREx mission \citep{Dore_2019}, the Nancy Grace Roman Space telescope \citep{Spergel_2015}, the Euclid mission \citep{euclid_presentaz}, and the LSST survey from the Vera Rubin Observatory \citep{LSST}, each expected to detect up to {\color{black} an order of} $10^5$ {\color{black}or more} voids. Large-scale surveys provide enough volume to measure voids over large areas of the sky while simultaneously observing low mass galaxies. Beyond increasing the statistical power of cosmic voids, we can now intensely study their {\color{black}properties}, such as by mapping their interior in detail.

Devoid of matter, comic voids are inherently extremely sensitive to the properties of diffuse components: they are a novel tool to constrain neutrino masses \citep{Paco_Lya, Massara_2015,Banerjee_2016, Sahlen_2018, Kreisch_2019,Schuster_2019,Zhang_2020,Bayer_2021}, modified gravity \citep[][]{Clampitt_2013,zivick_2015_grav,Cai_2015,Barreira_2015,Achitouv_2016,Achitouv_2019,Falck_2018,Sahlen_2018,Perico_2019,Contarini_2020,Alam_2021}, and dark energy \citep{Bos_2012,Spolyar_2013,Pisani_2015,pollina2016, Verza_2019}. The current understanding of our Universe sets the observed accelerated expansion of the Universe as one of the most puzzling mysteries for modern cosmology. The nature of dark energy, the component constituting $\sim68\%$ of the Universe and postulated to explain such accelerated expansion \citep{PlanckCollaboration_2018}, is still poorly understood. Since matter is missing in voids by definition, voids are {\color{black}regions in the Universe that become dark-energy dominated the earliest}, thus providing a window towards unraveling this dark mystery \citep{lee_DE_2009,Lavaux_2012}. The spherical property of void stacks in real space provides standard spheres for the \AP test \citep{alcock_paczynski,Ryden_1995,ryden_voids_rsd,Lavaux_2012,Biswas_2010}. Using stacked voids, it is also possible to model the redshift-space distortion (RSD) pattern around voids. The use of the void-galaxy cross-correlation function for the \AP test and RSDs is an area of intense data analysis-based activity in recent years \citep{Lavaux_2010}{\color{black}. Voids have provided results competitive to those from galaxies, and since voids do not suffer from non-linearities to the same extent as galaxies, there are indications that modeling could be easier for voids than for galaxies} \citep[e.g.][]{sutter2012_APtest,Sutter_2014VIDE,paz_rsd_2013,Hamaus_2016,Hawken_2016,Hamaus_2017,Hamaus_2020,Cai_2016,Cai_2017,Achitouv_2017,Correa_2019,Hawken_2020,Nadathur_2019,Nadathur_2020,Aubert_2020,Paillas_2021}. Aside from voids' use in cosmology, the study of galaxy properties in voids is also an active field of research \citep{Hoyle_2005,Patiri_2006,Kreckel_2012,Ricciardelli_2014,Habouzit_2020,Panchal_2020}.

Despite the wide use of voids in galaxy clustering analyses of survey data, simulation-based analyses of void capabilities with an extremely large set of simulations have only been performed using dark matter {\color{black}with neutrino} particles {\color{black} or dark matter particles alone} as tracers to find voids, and with a spherical void finder \citep{Bayer_2021}. To provide synergy with data analysis, a thorough investigation of voids extracted from the halo field and with a void finder preserving void shape and cosmic-web information is imperative. 

Relying on the large set of halo catalogues from the \quijote~simulations \citep{Villaescusa-Navarro_2020} and the shape-preserving void finder \vide~\citep{Sutter_2014VIDE}, we build the most extensive and realistic dataset of void catalogs ever released: the \gigantes\footnote{A reference to the giants in \textit{Don Quijote}'s world \citep{donquijote}.}~void catalogs suite, containing over 1 billion cosmic voids. The suite includes 15,000 \texttt{VIDE} fiducial cosmology void catalogs, as well as over 9,000 catalogs in non-fiducial cosmologies, spanning various values of the following cosmological parameters $\{\Omega_\mathrm{m},\Omega_\mathrm{b},h, n_s,\sigma_8,M_\nu,w\}$ and fully leveraging the \quijote~simulation suite, which covers redshifts $z=0.0,\,0.5,\,1.0,$ and $2$ in real- and redshift-space. 
The void finding relies on the popular public void finder \texttt{\texttt{VIDE}} \citep{Sutter_2014VIDE}, arguably the most used void finder, as testified by its use in a plethora of papers performing both simulation-based theoretical modelling and data analysis from modern surveys \citep[e.g.][]{Hamaus_2013,Hamaus_2014_universal,Sutter_2013,Sutter_2014VIDE,Chan_2014,Pisani_2014,Pisani_2015,Pollina_2017,Kreisch_2019,Verza_2019,cousinou2018,DES_2019a,DES_2019b,DES_2021}.
We provide further details on the \quijote~simulations and the \vide~void finder in \autoref{sec:quijote_vide}.

\begin{figure*}[t!]
\includegraphics[width=\linewidth]{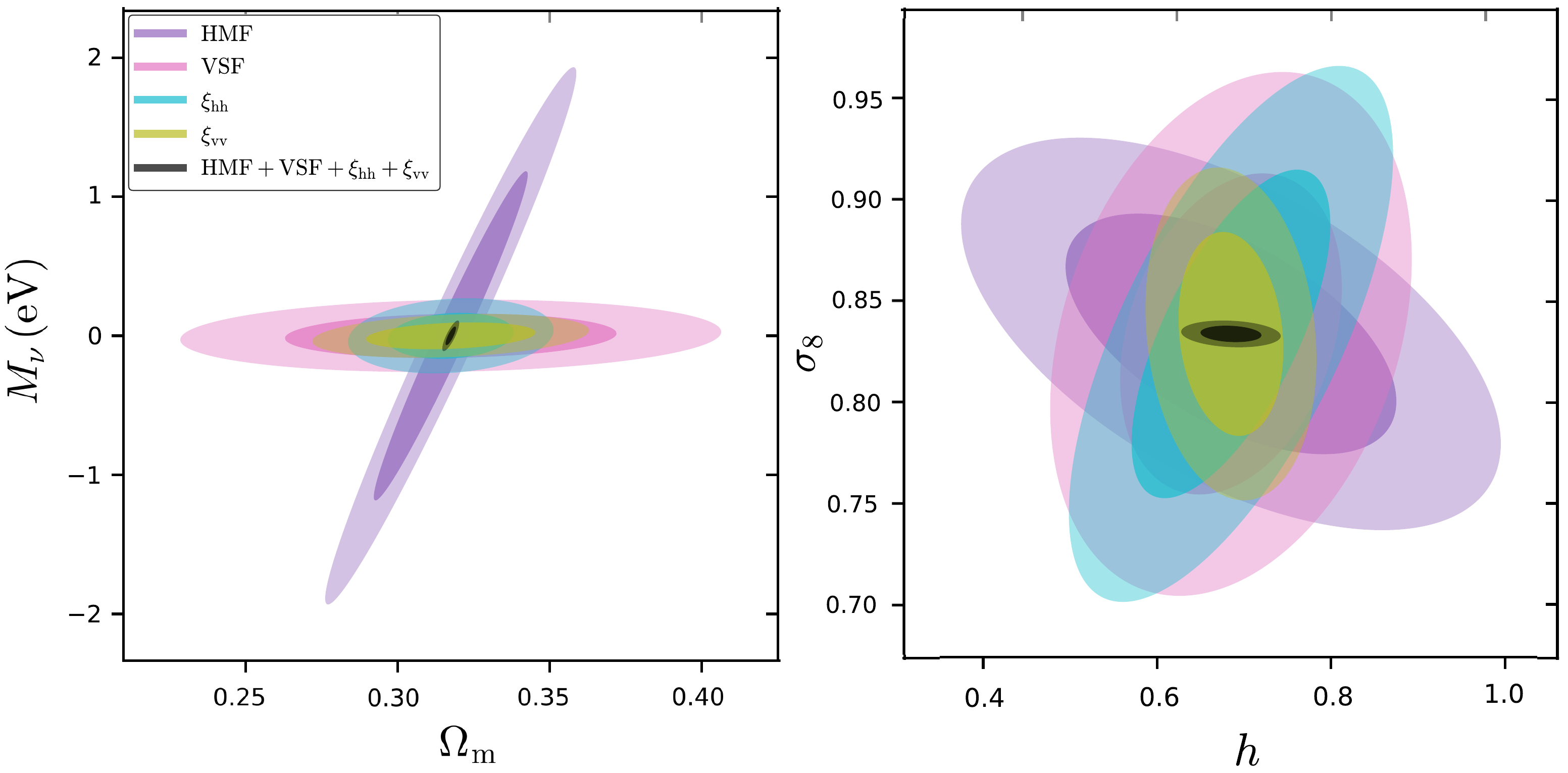}
\caption{Constraints on the sum of neutrino masses $M_\nu$, $\Omega_\mathrm{m}$, $\sigma_8$, and $h$ from the halo mass function (HMF), void size function (VSF), halo auto-correlation function ($\xi_{\rm hh}$), void auto-correlation function ($\xi_{\rm vv}$), and combined statistics {\color{black}for a volume of 1 $h^{-3}\mathrm{Gpc}^3$}. Void summary statistics provide strong constraints and orthogonal information to halo summary statistics.}
\label{fig:constraints}
\end{figure*}

Voids provide a set of summary statistics uniquely sensitive to the properties of the cosmological model. However, voids have been hotly debated as to whether or not they provide different information than the halo distribution they are found within. The main void summary statistics are the void-halo cross-correlation function $\xi_{\mathrm{vh}}$, equivalent to the void density profile, the void size function $n_{\mathrm{v}}$, a histogram of void sizes, and the void auto-correlation function $\xi_{\mathrm{vv}}$. 
With our suite of over 1 billion voids, we show in \autoref{fig:constraints} that void summary statistics carry novel and complementary information to halos, evidenced by the different degeneracies in the cosmological parameters {\color{black}for a volume of 1 $h^{-3}\mathrm{Gpc}^3$} (see Section \ref{subsec:additionalinfo} for details). 

Due to its unprecedented size and exploration of the power of different void summary statistics, including well known {\color{black}summary statistics} such as  $\xi_{\mathrm{vh}}$ and $n_{\mathrm{v}}$, but also recent {\color{black}summary statistics} such as $\xi_{\mathrm{vv}}$, and its power in accurately estimating covariance matrices, the \gigantes~voids dataset constitutes a powerful benchmark for machine learning applications. Its massive number of voids coupled with a large number of realizations permits the extraction of a strong void signal, information that is usually obscured by noise due to a low number of voids. As such, the \gigantes~suite opens the realm of machine learning exploration for voids, a new avenue for these objects. With such a wealth of data for voids, the \gigantes~suite provides a testbed not only for developing new cosmological theories, but also for developing novel machine learning techniques. 
Given the nature of voids, they can be analyzed with traditional techniques, such as CNNs (when projected into 2D or 3D grids) albeit with the caveats mentioned earlier, or GNNs (when exploiting the full sparsity of the data). Their symmetries, need for interpretation, and need for uncertainty estimation, however, pose questions ideally suited {\color{black}for} the development of new machine learning techniques.

This paper is organized as follows{\color{black}, and all results are presented for a volume of 1 $h^{-3}\mathrm{Gpc}^3$}. In Section \ref{sec:gigantes} we present the \gigantes~dataset as well as the standard three void summary statistics. In  Section \ref{subsec:additionalinfo} we address the question on whether voids carry additional information {\color{black}on cosmological parameters} with a Fisher matrix forecast {\color{black} in real space at $z=0$} exploring the constraining power of voids obtained from the halo field with respect to traditional probes. {\color{black}We leave for future work an analysis in redshift space.} In Section \ref{sec:info} we study where the information comes from, with a computation of the correlation between and among void and halo summary statistics, a Fisher forecast {\color{black} in real space} exploring the combined power of voids and halos, and a comparison of the power of void shape in constraints to a simplistic spherical void finder. {\color{black} This paper focuses on illuminating the full information contained in these probes rather than exploring models for such probes.} In Section \ref{sec:LFI} we show a machine learning application by training a neural network to perform likelihood-free inference from the void size function. We draw the main conclusions of this work in Section \ref{sec:conclusion}.

\section{\gigantes} 
\label{sec:gigantes}

This section describes the void catalogs and the primary void summary statistics. 

\begin{figure*}[t!]
\includegraphics[width=\linewidth]{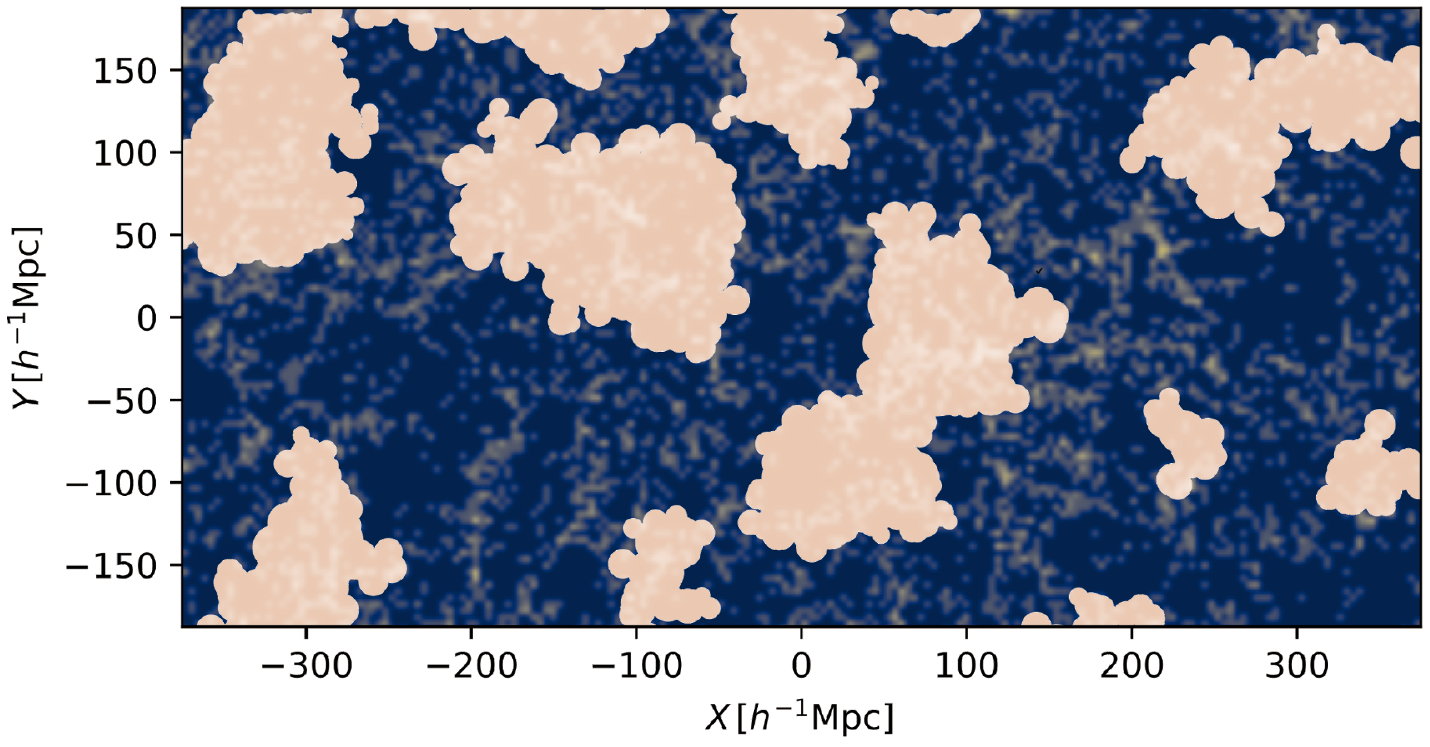}
\caption{Several randomly selected voids are projected on the halo density field in orange, illustrating a range of sizes and shapes. The background shows the halo density field over a region of $375\times750\times250~(h^{-1}{\rm Mpc})^3$. Void shapes are approximated by treating Voronoi cells as circles and projecting them on the halo density field. Despite the presence of projection effects, this 2D representation of voids in the slice illustrates the complex substructure of voids and the rich variety {\color{black}in their} shapes.}
\label{fig:videvoid}
\end{figure*}

\subsection{\gigantes~catalogs}
We build \texttt{VIDE} void catalogs from all the \quijote~simulations at different redshifts. 
A summary of all catalogs available is shown in \autoref{table:voidcatalogs}. \gigantes~ includes catalogs for the fiducial cosmology, catalogs varying single parameters for Fisher matrix calculations, and catalogs sampling multiple parameters at a time from a latin hypercube. We provide catalogs at different redshifts (0.0, 0.5, 1.0 and 2.0) as well as for real-space and redshift-space. Overall, the \gigantes~suite provides more than 1 billion voids over thousands of different cosmological models. The \gigantes~dataset will be publicly released upon acceptance of the paper. 

Each void catalog provides a set of files as described in \cite{Sutter_2014VIDE}, giving for each void a set of properties, including:
\begin{enumerate}
    \item the void center position (RA, Dec, z for observations; x, y, z for simulations),
    \item the void radius (see \autoref{sec:quijote_vide}),
    \item the void volume,
    \item the void density contrast (as defined by \texttt{ZOBOV} and thus used by \texttt{VIDE}, given by the ratio of the minimum density along the ridge of the void versus the minimum density in the void),
    \item the number of particles of each void (halos in our case),
    \item the central density \citep[density in a sphere of $0.25R_\mathrm{v}$, see][]{Sutter_2014VIDE},
    \item the void ellipticity,
    \item and the void hierarchy. 
\end{enumerate}
Since \texttt{VIDE} voids are found taking into account the hierarchical feature of the cosmic web, the void finder outputs all voids (parents) and their sub-voids (children): for each void, the catalog includes the ID of the void parent (if applicable), the tree level, and the number of children.

\begin{center}
\begin{table}[!t]
\hskip-1.0cm\begin{tabular}{lllll}
\hline
\multicolumn{1}{|c|}{Cosmology}        & \multicolumn{1}{c|}{Redshifts} & \multicolumn{1}{c|}{Realizations (real space)} & \multicolumn{1}{c|}{Realizations (redshift space)} & \multicolumn{1}{c|}{Total void \#} \\ \hline
\multicolumn{1}{|c|}{Fiducial}         & \multicolumn{1}{c|}{[0.0,0.5,1.0,2.0]}          & \multicolumn{1}{c|}{15,000 \textit{(per z-bin)}}             & \multicolumn{1}{c|}{{15,000\textit{(per z-bin)}} }                    & \multicolumn{1}{c|}{$\sim 400,000,000$}             \\ \hline
\multicolumn{1}{|c|}{$\pm$ for Fisher} & \multicolumn{1}{c|}{[0.0,0.5,1.0,2.0]}          & \multicolumn{1}{c|}{$500\times 9$ \textit{(per z-bin \& $\pm$ parameter)}}             & \multicolumn{1}{c|}{$500\times 9$ \textit{(per z-bin \& $\pm$ parameter)}}                    & \multicolumn{1}{c|}{$\sim 300,000,000$}             \\ \hline
\multicolumn{1}{|l|}{ Latin Hypercube}                 & \multicolumn{1}{l|}{[0.0,0.5,1.0,2.0]}          & \multicolumn{1}{c|}{2,000\textit{(per z-bin \& resolution)}}             & \multicolumn{1}{c|}{2,000\textit{(per z-bin \& resolution)}}                    & \multicolumn{1}{c|}{$\sim 350,000,000$}             \\ \hline

\end{tabular}
\caption{Summary of the \gigantes ~void catalogs. The fiducial cosmology used in the simulations is: $\Omega_\mathrm{m}=0.3175$, $\Omega_\mathrm{b}=0.049$, $h=0.6711$, $n_s=0.9624$, $\sigma_8=0.834$, $M_\nu=0.0\,\mathrm{eV}$, $w=-1$. Void catalogs for Fisher analyses are derived from simulations with a slight variation for each parameter.  
}\label{table:voidcatalogs}
\end{table}
\end{center}

\subsection{Main void summary statistics}\label{sec:observables}

The void finder provides void centers and radii, which in themselves yield two different summary statistics to be used in cosmological data analyses: the void auto-correlation function, and the void size function. Combining void positions with halo positions defines a third summary statistic: the void-halo {\color{black}cross-}correlation function.

\subsubsection{The void-halo cross-correlation function $\xi_\mathrm{vh}$}

By knowing the position of void centers and halos, it is possible to compute the probability of finding a halo at a certain distance from the void center, known as the void-halo cross-correlation function, $\xi_\mathrm{vh}$, or the void density profile {\color{black} found in the halo field}. This probability will be low in the center of the void (it is not very likely to find halos, or galaxies, close to the void center), and high at the over-dense rim enclosing the void{\color{black}, which is coincident with the average void radius}. It is usually computed considering a stack of many voids. While the estimation of the void-halo cross-correlation function from data usually relies on estimators, it can be shown that $\xi_\mathrm{vh}$ is equivalent to the void density profile \citep[see Eq. (2.7) and (2.8) of][]{hamaus_rsd_2015}:
\begin{equation}
    \xi_\mathrm{vh}(r)=\frac{\rho_{\mathrm{vh}}(r)-\bar{\rho}_\mathrm{h}}{\bar{\rho}_\mathrm{h}}
\end{equation}
where $r$ is the distance between the halos and the void center, $\rho_{\mathrm{vh}}$ is density of halos in shells around the void center, and $\bar{\rho}_\mathrm{h}$ is the average tracer density of the dataset. 

The void density profile is sensitive to cosmological parameters in both real-space and redshift-space. The peak height and location of the 1D real space profile, denoting the void wall's rim {\color{black}and average void radius}, responds to changes in cosmological parameters. In a homogeneous and isotropic universe void density profiles do not possess a preferred direction. Thus, in real space, the 2D density profile will be spherically symmetric. In redshift space, however, the spherical stacked 2D void profile will be distorted due to the \AP effect, which introduces cosmology-dependent geometrical distortions along the line-of-sight \citep{alcock_paczynski}. The 2D redshift space profile is also distorted due to RSDs. The magnitude of these effects depend on the cosmological model, so the observed 2D void density profile can be used to constrain cosmology. 

The void density profile has been used to successfully measure the \AP effect by using voids as standard spheres \citep[e.g.][]{sutter2012_APtest} {\color{black}and} simultaneously modeling RSDs \citep[e.g.][]{Hamaus_2016,Hamaus_2020}. The profile provides the most stringent {\color{black}constraint on the Universe's expansion history at $z=0.5$ to date from voids}: a calibration-free measurement from voids only, with a $16.8\%$ precision level measurement of the growth rate of structure and an independent measurement of the matter content of the Universe yielding $\Omega_{\mathrm{m}}=0.312 \pm 0.020$ \citep{Hamaus_2020}.

\begin{figure}[t!]
\begin{minipage}[t]{0.5\linewidth}
\includegraphics[width=0.99\linewidth]{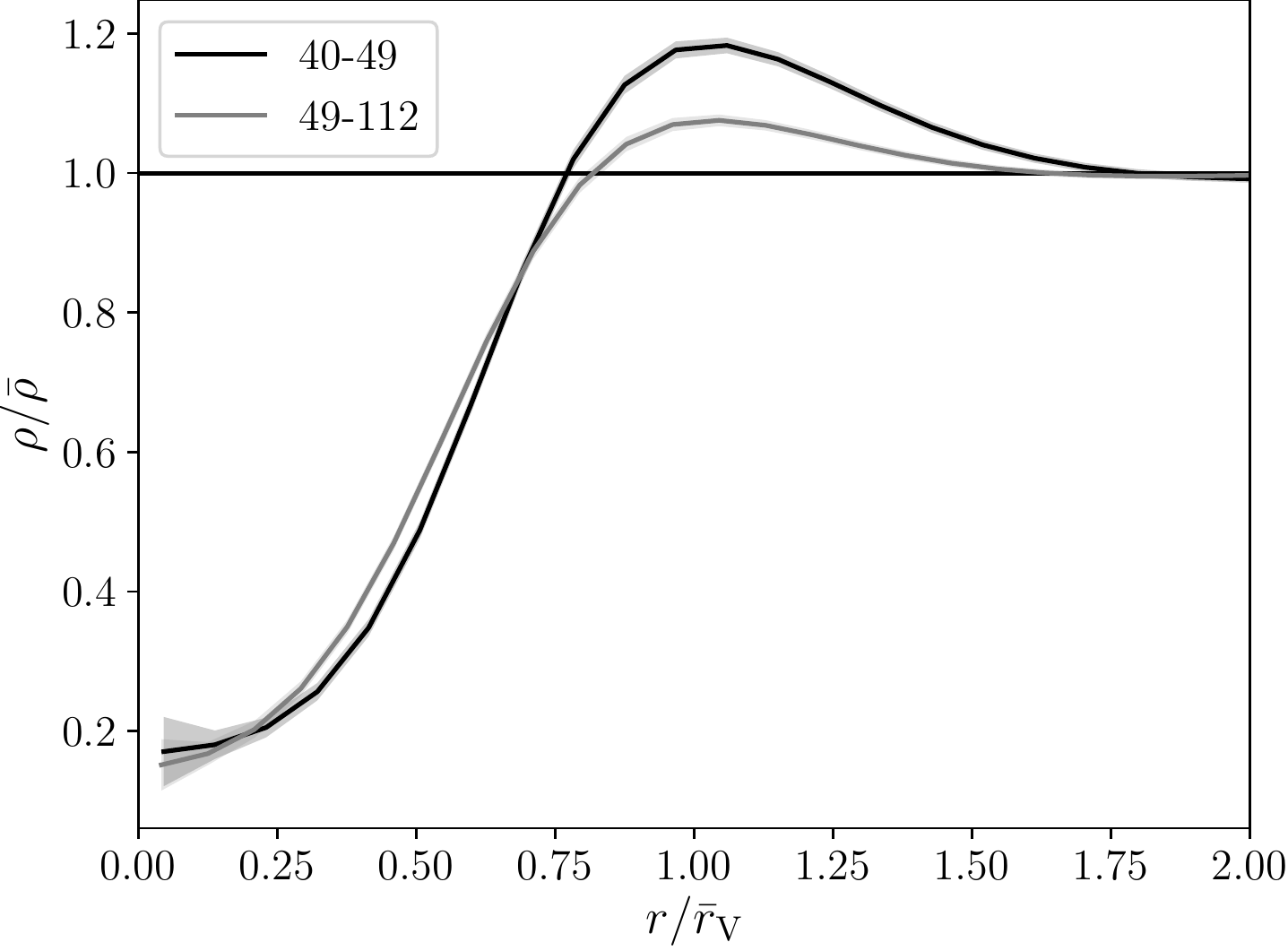}
\caption{Void density profiles for 2 different radii bins in $h^{-1}\,\mathrm{Mpc}$ at $z=0$. The void wall rim occurs at the peak of the density profile{\color{black}, which is also coincident with the average void radius}.}
\label{fig:Xvh}
\end{minipage}
\hspace{0.1cm}
\begin{minipage}[t]{0.5\linewidth} 
\includegraphics[width=0.99\linewidth]{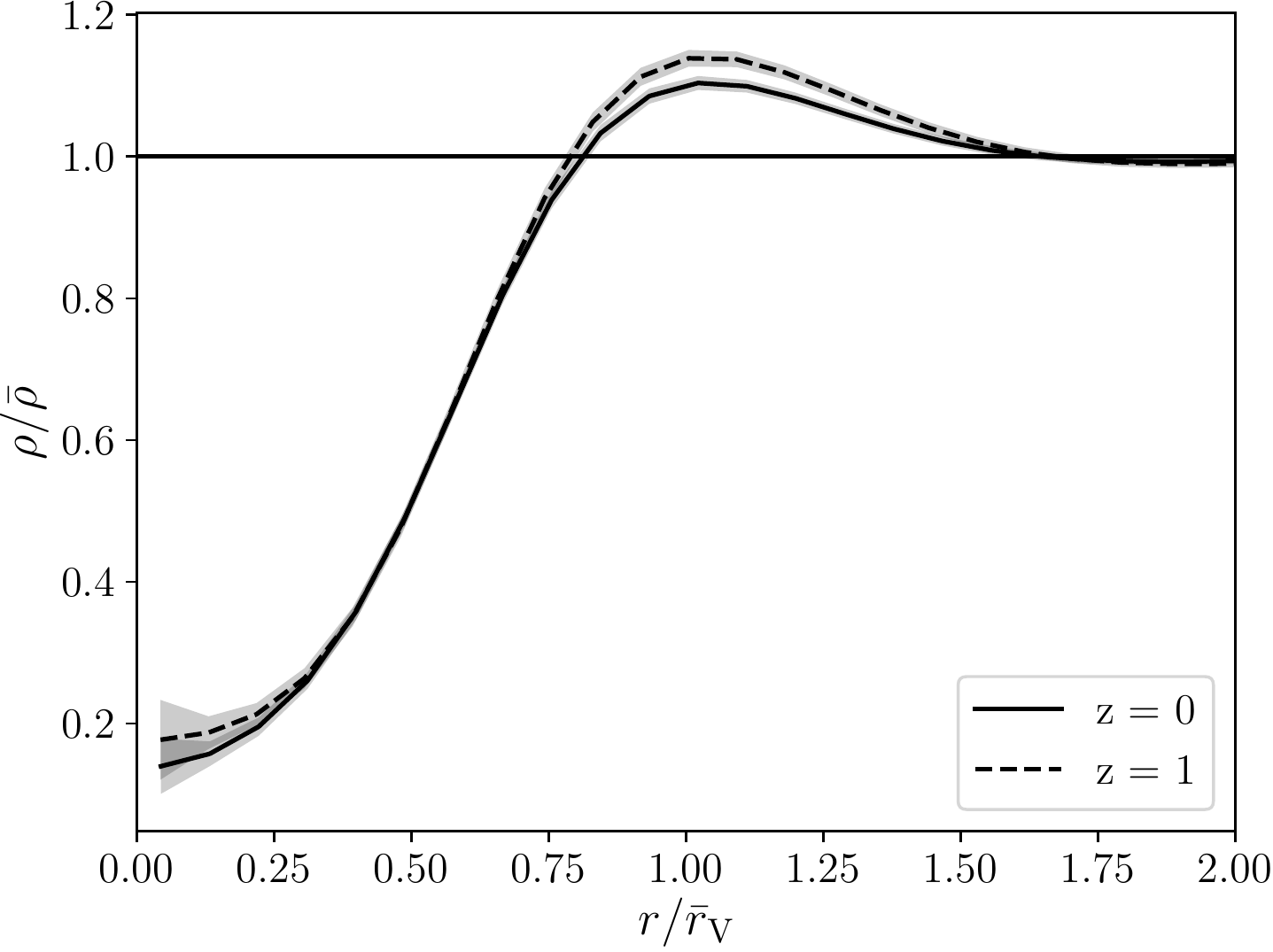}
\caption{Void density profiles for $z=0$ and $z=1$ with void radii spanning 50-70 $h^{-1}\,\mathrm{Mpc}$ for each.}
\label{fig:Xvh_Z}
\end{minipage}
\end{figure}

\autoref{fig:Xvh} shows two density profiles measured from the \gigantes~void catalogs for voids with radii 40-49  $h^{-1}\,\mathrm{Mpc}$ and 49-112 $h^{-1}\,\mathrm{Mpc}$ at $z=0$. Distance is normalized by the average void radius for the denoted radius bin. Larger voids have a less pronounced void wall rim than smaller voids due to the fact that larger voids have evolved more than smaller voids at a fixed redshift {\color{black}and because we find voids in the halo field rather than in the dark matter field}. \autoref{fig:Xvh_Z} shows profiles from two different redshift bins ($z=0$ and $z=1$) for voids 50-70  $h^{-1}\,\mathrm{Mpc}$ in radius\footnote{Errors are propagated for the average correlation function based on the individual errors estimated by \vide, which are based on scatter in the bin average.}. For the same radius bin, voids at present are more evolved than voids at $z=1$ since they are, on average, older. Thus, voids at $z=0$ are more evolved and so have a lower void wall rim than voids at $z=1$. 

\subsubsection{The void size function}

The void size function measures the number of voids as a function of their radius, also known as the void abundance, and is particularly sensitive to cosmological parameters, such as the dark energy equation of state \citep{Pisani_2015,Verza_2019}, modified gravity \citep{Contarini_2020} and massive neutrinos \citep{Kreisch_2019,Sahlen_2018}. The theoretical modeling of the void size function is an area of intense activity \citep[e.g.][]{Sheth_2003,Jennings_2013,Pisani_2015,Sahlen_2016,contarini2019,Verza_2019}, and will provide stringent constraints when applied to the next generation of surveys \citep[e.g.][]{Pisani_2015,Dore_2019}. The main theoretical model to be considered for the abundance is the Sheth and van de Weygaert model \citep{Sheth_2003}, an excursion-set based prediction of void numbers of different sizes, later extended to consider the volume conservation of voids, since small voids merge to give larger voids \citep{Jennings_2013}. This model has been shown to reproduce simulation results in different cosmologies where either the dark energy equation of state \citep{Pisani_2015,Verza_2019} is varied or modified gravity is considered \citep{Contarini_2020}. Since the model {\color{black}assumes dark matter voids and spherical symmetry}, current work focuses on improving the match of predictions with measurements of observed voids in mocks \citep[e.g. by including tracer bias, see][]{Pollina_2018,contarini2019,Contarini_2020}. 

In this work we do not attempt to model the measured void abundance with theoretical models, but we instead choose the approach of relying on the measured values of void numbers from the \gigantes~dataset to assess its constraining power for cosmology. In \autoref{fig:constraints} we show a Fisher forecast {\color{black}over a volume of 1 $h^{-3}\mathrm{Gpc}^3$} for constraints on a few cosmological parameters from the void size function for radii $6-60\,h^{-1}\mathrm{Mpc}$. The void size function puts superior constraints on $\sum m_\nu$, for example, compared to the halo mass function (errors of $0.21\,\mathrm{eV}$ and $1.56\,\mathrm{eV}$, respectively), and is competitive with the halo auto-correlation function (error of $0.22\,\mathrm{eV}$). See \autoref{sec:error} and \autoref{sec:full_constraints} for forecasted constraints on additional cosmological parameters. {\color{black} In \autoref{sec:error} we also illustrate the stability of the Fisher uncertainties by varying the number of simulations used to calculate the uncertainties and illustrating that the change in the value of the uncertainties is less than the change in the number of simulations.}

\begin{figure}[t!]
\begin{minipage}[t]{0.5\linewidth}
    \centering
    \includegraphics[width=0.99\linewidth]{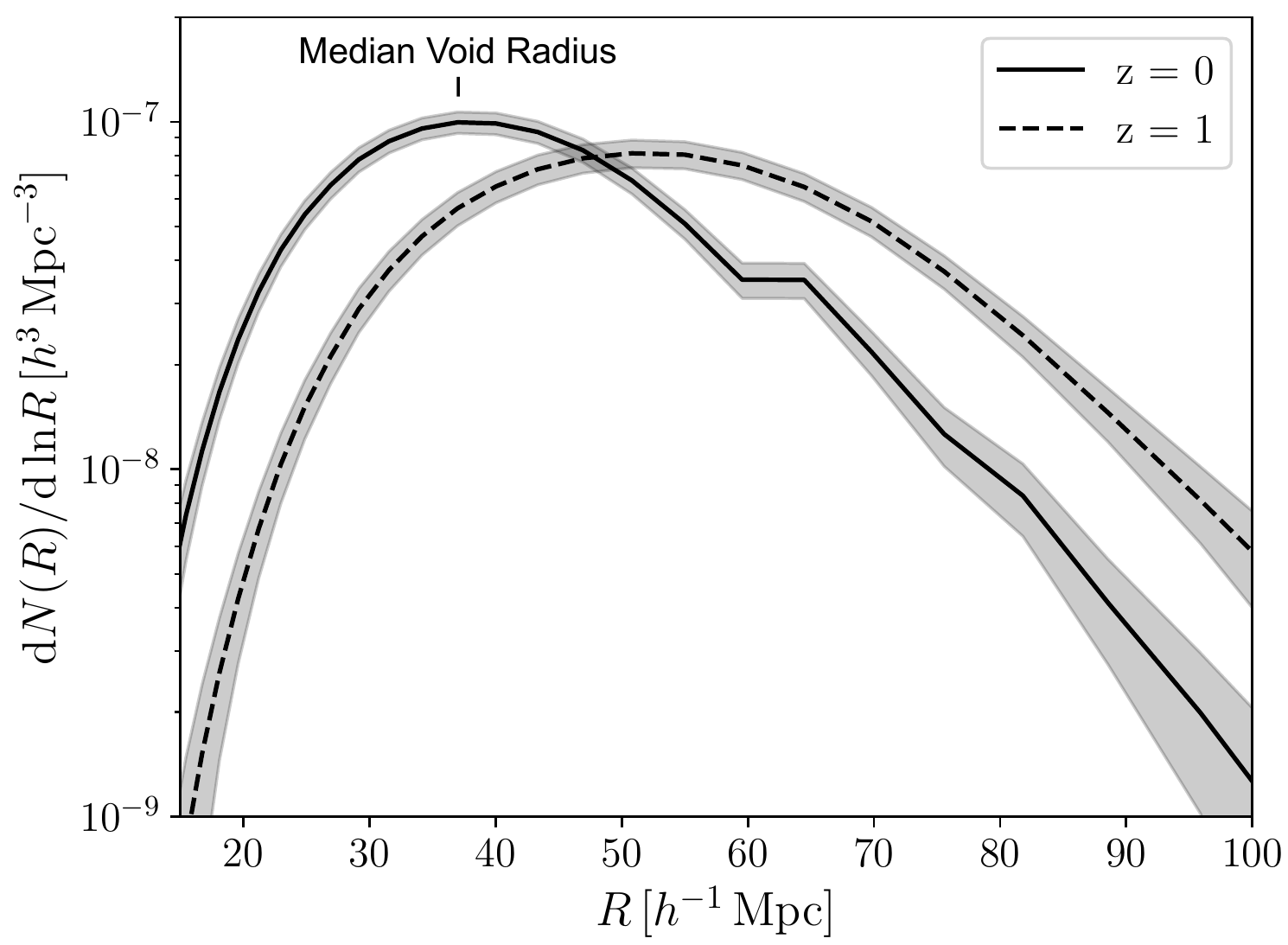}
    \caption{Void size function, or void abundance, from the \gigantes~dataset for both $z=0$ and $z=1$. The peak of the abundance corresponds to the {\color{black}median} void radius.}
    \label{fig:abundance}
\end{minipage}
\hspace{0.1cm}
\begin{minipage}[t]{0.5\linewidth} 
    \centering
    \includegraphics[width=0.99\linewidth]{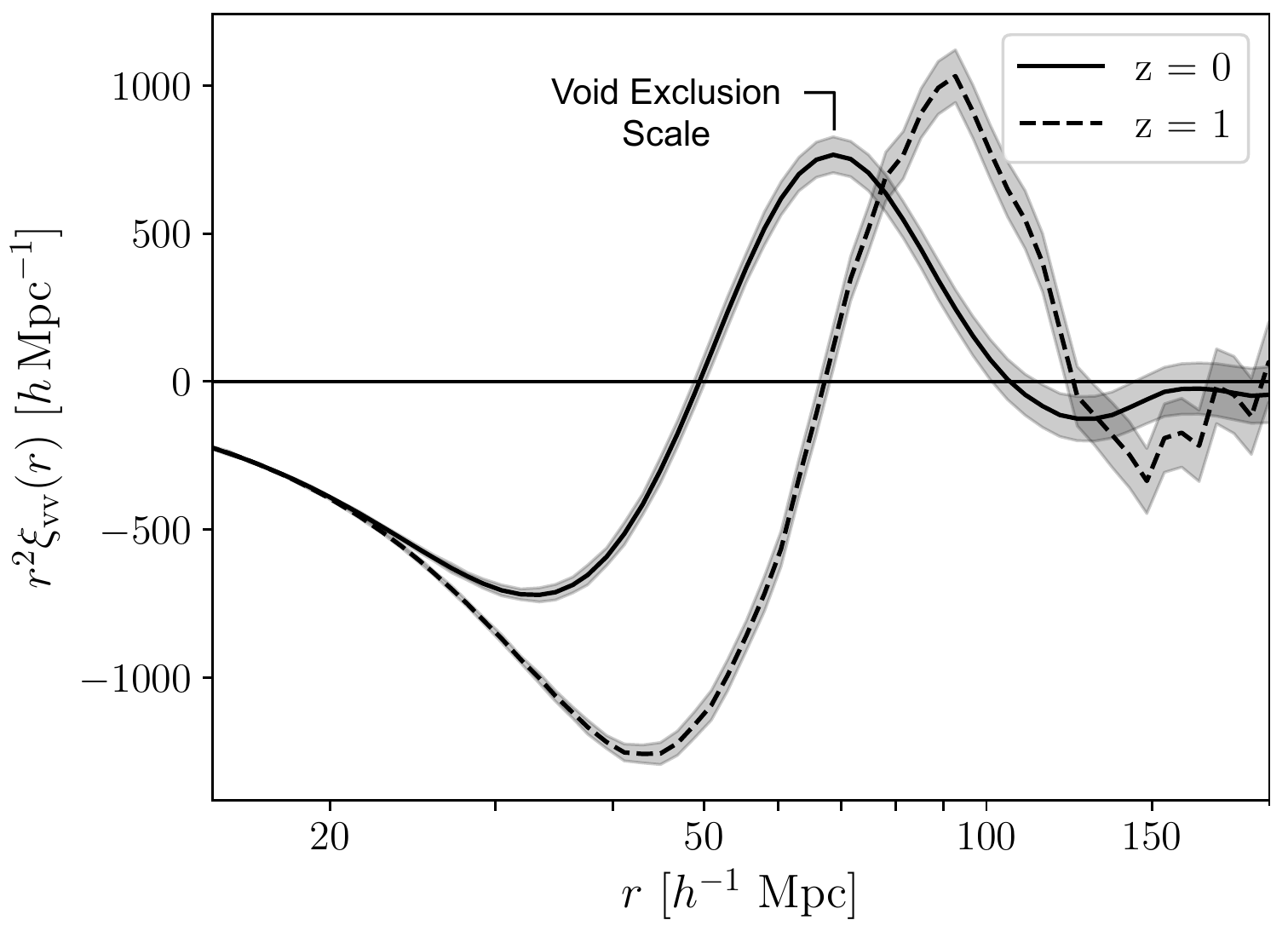}
    \caption{Void auto-correlation function $\xi_\mathrm{vv}$ from \gigantes~voids, for both $z=0$ and $z=1$. The void exclusion scale is coincident with the peak of the void auto-correlation function {\color{black}at $\sim 2\tilde{R}_{\rm V}$, where $\tilde{R}_{\rm V}$ is the median void radius}.}
    \label{fig:Xvv}
\end{minipage}
\end{figure}

\autoref{fig:abundance} shows the abundance of voids from the \gigantes~dataset at $z=0$ and at $z=1$. As described below in Section \ref{subsec:xivv}, voids are larger in the latter case. We show the void size function for radii ranging from $15$ to $100\,h^{-1}\,\mathrm{Mpc}$. The {\color{black}median} void radius is $35\,h^{-1}\mathrm{Mpc}$ {\color{black}for $z=0$}. Voids of lower size are less abundant due to the low simulation resolution at small scales (the mean tracer separation is $\sim 13\,h^{-1}\,\mathrm{Mpc}$ at $z=0$ and $\sim 17\,h^{-1}\,\mathrm{Mpc}$ at $z=1$). Nevertheless, small voids improve cosmological constraints, pointing to the constraining power of the void hierarchy, which will be further explored in future work. The large volume and, thus, strong void statistics in the \gigantes~suite promotes a high signal-to-noise ratio for the void size function's response to cosmological parameters.

\subsubsection{The void auto-correlation function $\xi_\mathrm{vv}$}
\label{subsec:xivv}

We can compute the probability of finding a void center at a given distance from a randomly selected void in our void catalog, yielding the void auto-correlation function. Since the center of a void is defined as the volume-weighted barycenter of the Voronoi cells composing the void, it carries information about the whole structure of the void. This allows us to uncover a new sensitivity of the large-scale structure of the Universe to cosmological parameters (see Section \ref{subsec:additionalinfo}, Section \ref{sec:info}, and Section \ref{sec:LFI}). The void auto-correlation function requires a great number of voids to overcome shot-noise, and it has so far been measured from data only at low redshift, over a volume of $0.6\, (h^{-1}\mathrm{Gpc})^3$ \citep{Clampitt_2016}, and considering relatively small effective void radii  ($<30 h^{-1}\mathrm{Mpc}$).

Given its limited application to data, the void auto-correlation function has not yet been used to derive constraints on parameters of the cosmological model. It has been investigated in simulations \citep{Chan_2014,Hamaus_2014} and shows a strong sensitivity to massive neutrinos \citep{Kreisch_2019}. We further illustrate this sensitivity to neutrinos {\color{black}for a volume of 1 $h^{-3}\mathrm{Gpc}^3$} in \autoref{fig:constraints}, in which the void auto-correlation function puts the strongest forecasted constraints on $\sum m_\nu$ of $0.13\,\mathrm{eV}$. {\color{black}One reason for $\xi_{\rm vv}$'s strong constraints on the parameters could be that it contains a strong, clear feature at the void exclusion scale. Such strong features are useful for measuring distances, like the BAO, probing the background cosmology \citep[see also][]{Hamaus_2014}. Further, void bias can be large \citep[see e.g.][]{Chan_2014,Schuster_2019}, and this boosts the large-scale correlation function amplitude which can provide a better signal-to-noise ratio to measure the large-scale power spectrum, which depends on $\sigma_8$ and $n_{\rm s}$.} See \autoref{sec:error} and \autoref{sec:full_constraints} for further forecasted constraints on cosmological parameters. 

With the advent of upcoming surveys, such as the DESI experiment \citep{DESI2016}, the Roman Space Telescope \citep{Spergel_2015}, the Euclid ESA satellite \citep{euclid_presentaz}, the SPHEREx mission \citep{Dore_2019}, the Rubin Observatory \citep{LSST}, and the PFS Subaru survey \citep{pfs2016}, hundreds of thousands of voids will be observed \citep{Pisani_2019}. The void auto-correlation function will then reach the realm of observable quantities that can be measured from data with high significance, and used to constrain cosmological parameters.

We show the void auto-correlation function measured from the \gigantes~data set in \autoref{fig:Xvv} at $z=0$ and $z=1$. 
The void auto-correlation function has a maximum at twice the mean void radius of the sample. This corresponds to the void exclusion scale \citep{Hamaus_2013}, which is termed as such because, on average, void walls touch at twice the mean void radius since voids cannot overlap. This makes it likely to have two void centers at a distance corresponding to twice the mean void radius of the sample. Voids on average cannot be separated by smaller scales because then they will overlap. However, there is a distribution of void sizes (i.e. the void size function), and smaller voids are allowed to be separated by equivalently smaller scales. Thus, the probability of two voids being separated smoothly decreases as scales become smaller, since there are fewer small voids.

At higher redshift, the observed average radius of \gigantes~voids is larger. The Universe is more dominated by dark energy at $z=0$ than it was at $z=1$. 
Large, isolated voids in the dark matter field {\color{black}mostly expand} in their life time. Large voids in the dark matter field at $z=1$ most likely continue to expand through $z=0$. Here, however, we consider voids found in the halo field. The number density of halos of a lower fixed mass of, e.g., $10^{13}\,\mathrm{M}_\odot$, is higher at present. This high number density of small halos results in, on average, more small voids at present than at $z=1$. {\color{black}This is also why the peak of the void density profile at $z=0$ is shallower than at $z=1$, whereas in dark matter simulations the opposite occurs \citep{Hamaus_2014_universal}.}
Thus, the average void size is larger at $z=1$ in the \gigantes~void catalogs suite. We also note that at $z=1$ there are less voids, so the $\xi_\mathrm{vv}$ is noisier.
The theoretical modelling of the void auto-correlation function is an area that will deserve further attention in coming years, given its promising constraining power (see Section \ref{subsec:additionalinfo}). 

\section{Do voids carry additional information? Yes.} \label{subsec:additionalinfo}

The question of whether voids, defined by relying on the halo (or galaxy) distribution, carry additional cosmological information with respect to traditional tools based on the halo (or galaxy) distribution, such as the two-point correlation function or the halo-mass function, has been a debated topic. 

From a theoretical perspective, there have been indications to believe voids contain higher order information. As extended 3-D objects, voids must be defined by at least 4 non-planar halos. The fact that voids are defined by many halos could hint to the possibility of accessing higher-order information, but this is not in principle guaranteed: it is possible to find extended 3-D objects, and, thus, voids, in a Gaussian random field \citep[or a {\color{black}Poisson distribution of tracers}, see][]{cousinou2018}. 
Nevertheless, the universe at present is not a Gaussian random field, and certain summary statistics associated to voids, like the void size function, have been expected to contain information beyond the 2-pt correlation function. As 1-point functions, amplitude and shape are expected to depend on all n-point functions. The probability of a random volume being a void can be written in terms of $n$-point correlations \citep{fry1986,fry2013}:
\begin{equation}
    P_0 = {\rm exp}\left[\sum_{n=1}^\infty \frac{\left(-1\right)^n}{n!} \bar{N}^{n} \xi_n \right],
\end{equation}
where $\bar{N}$ is the mean tracer (i.e. halo or galaxy) count. {\color{black}Further, we are using voids found in the halo field, which already probes higher $k$. Another possible explanation for why voids can capture non-linear information is that an individual void is built from halos separated by distances approximately less than or equal to its diameter, highlighting scales that are quasi-linear and approaching non-linear.}

This paper has the tools necessary to properly investigate this question, relying on a comprehensive Fisher analysis.
We investigate the power of the void statistics mentioned in Section \ref{sec:observables} with respect to two traditional statistics defined from the halo field: the halo mass function and the halo auto-correlation function. The halo mass function measures the abundance of halos of a given mass, while the halo auto-correlation function gives the probability that, taken a random halo, there is another halo at a distance $r$ from the first halo. 

\autoref{fig:constraints} shows constraints from different void summary statistics, as well as for the halo mass function and the halo auto-correlation function, {\color{black} in real space} {\color{black}over a volume of 1 $h^{-3}\mathrm{Gpc}^3$} for selected parameters: the sum of neutrino masses $\sum m_\nu$, the matter content of the Universe $\Omega_\mathrm{m}$, the reduced Hubble constant $h$, and {\color{black}the amplitude of linear density fluctuations inside spheres of radius 8 $h^{-1}\mathrm{Mpc}$} $\sigma_8$. {\color{black}See Section 4.1 in \citet{Villaescusa-Navarro_2020} for detailed information on how the Fisher analysis is performed.} \cite{Bayer_2021} analysed the power of voids found in the dark matter distribution, but here we use the realistic set of \gigantes~void catalogs drawn from the halo distribution to show that voids observed by galaxy surveys hold the power to strongly reduce the available parameter space for cosmological parameters. 

From this figure we clearly see that the void size function constraints have a very different orientation with respect to constraints from the halo mass function and the halo auto-correlation function. The void auto-correlation function provides even stronger constraining power as well as access to unique information, as evidenced from its orientation with respect to the other three summary statistics. The void size function, in particular, is almost fully orthogonal to the halo mass function, illustrating it provides access to information beyond that accessible from the halo statistics. Further, the orientation of the void and halo auto-correlations differ, and for some parameter planes, such as the $\sigma_8-h$ plane, the halo mass function, void size function, halo auto-correlation function, and void auto-correlation function are almost all in different orientations.

The combination of all the mentioned summary statistics considerably tightens final constraints, demonstrating that voids carry complementary information to that of halos. 
See \autoref{sec:full_constraints} for the constraining power for other parameters.

\section{Where the information comes from} 
\label{sec:info}

Aside from showing that voids provide additional information, the large \gigantes~dataset allows us to estimate how correlated the different summary statistics are and where the information {\color{black}on cosmological parameters} comes from. The \gigantes~dataset contains void catalogs similar to those expected from future surveys such as PFS, since they have a similar number density to the \quijote~simulations, although different halo biases. Upcoming surveys such as Euclid and the Roman Space Telescope will have a higher number density than the \quijote~simulations, and so are expected to provide even better constraints from voids than what is illustrated here. See \autoref{sec:quijote_vide} for details on comparing the \quijote~simulations to upcoming surveys.

\subsection{Summary Statistic Correlation} \label{subsec:corr}

\begin{figure*}[t]
\includegraphics[width=\linewidth]{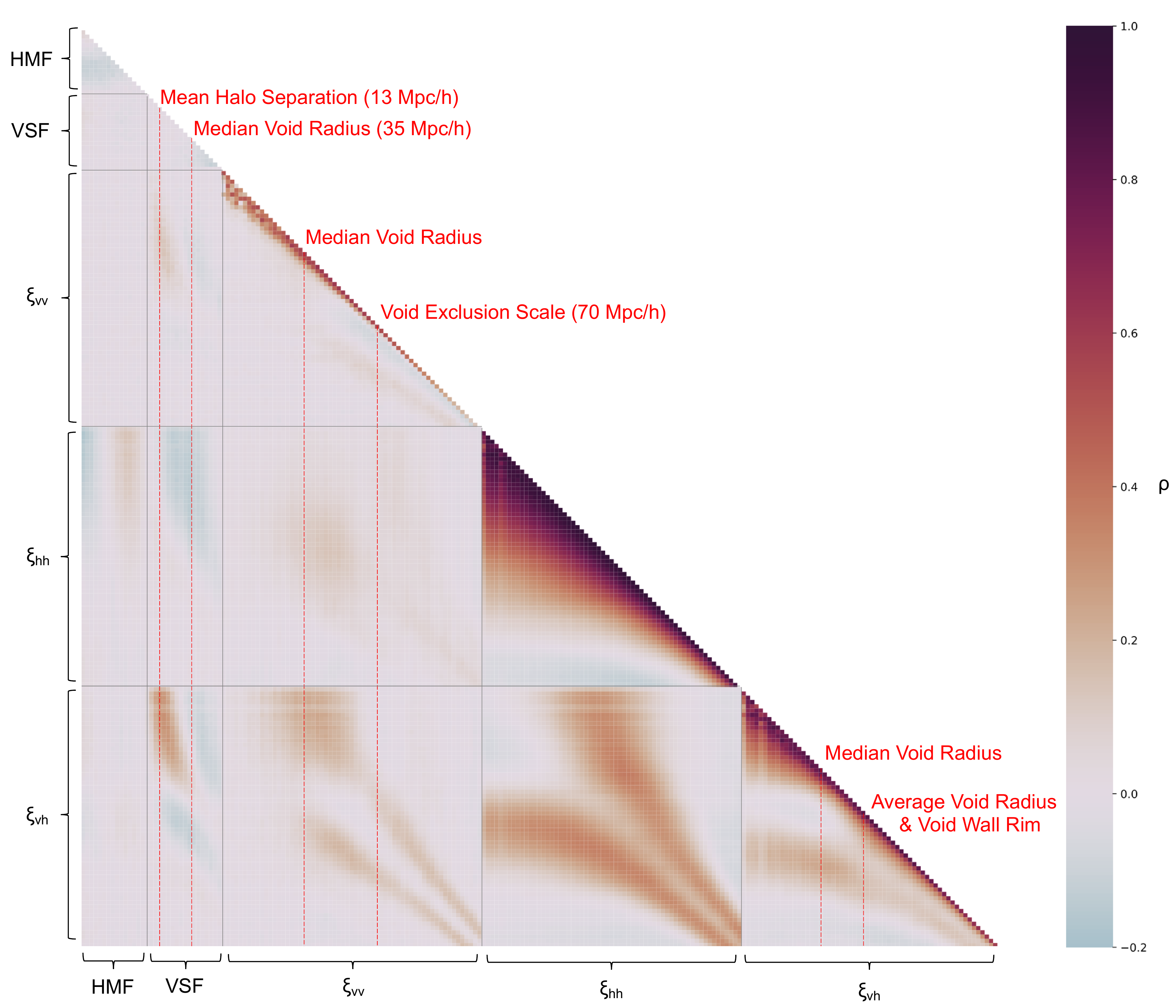}
\caption{Correlation for the halo mass function (first 15 bins), the void size function (following 18 bins), the void auto-correlation function (following 61 bins), the halo auto-correlation function (following 61 bins), and the void-halo cross-correlation function (last 61 bins). All the correlation functions span from 15 to 200 $h^{-1}\mathrm{Mpc}$ in logspace. The VSF goes from 6 to 60 $h^{-1}\mathrm{Mpc}$ in increments of 3 $h^{-1}\mathrm{Mpc}$.}
\label{fig:corr}
\end{figure*}

An important point to address is how the different void statistics are correlated, and if information from different bins is correlated. In \autoref{fig:corr} we show the correlation matrix between and within the halo mass function (first 15 bins), the void size function (following 18 bins), the void auto-correlation function (following 61 bins), the halo auto-correlation function (following 61 bins), and the void-halo cross-correlation function (last 61 bins). We utilize the 15,000 fiducial simulations at $z=0$ and compute the standard Pearson correlation coefficient for the different summary statistics:
\begin{equation}
    \rho(x,y)=\frac{\mathrm{Cov}\left(x,y\right)}{\sigma_x \sigma_y}.
\end{equation}
The correlations are computed {\color{black}for a volume of 1 $h^{-3}\mathrm{Gpc}^3$} for scales utilized in the Fisher forecast, namely 15 to 200 $h^{-1}{\rm Mpc}$ {\color{black}in log space} for the correlation functions and radii 6-60 $h^{-1}{\rm Mpc}$ for the void size function. {\color{black}Bins for the HMF are used from \citet{Bayer_2021}. This, however, does not make a statement on whether or not the parameter constraints are independent, as 0 correlation does not imply parameter independence. In \autoref{sec:error} we also illustrate the stability of the covariance by varying the number of simulations used to calculate the covariance and illustrating that the change in the value of the uncertainties is less than the change in the number of simulations.}

The void size function and the halo mass function each have relatively uncorrelated bins. A few bins at small distances (below the $\color{black}median$ radius of 35 $h^{-1}{\rm Mpc}$) in the void auto-correlation function are correlated, indicating some coupling at those scales, while larger scales are decoupled. The halo auto-correlation function is more strongly correlated over all scales, while the void-halo correlation function shows a slightly higher correlation at low radii (that is, below the average void radius) than the void auto-correlation function. {\color{black} We note that scales roughly below $20\,h^{-1}\mathrm{Mpc}$ in the correlation matrix show some scatter. This may be physical due to \quijote~being well resolved below such scales and the covariance being stable (see \autoref{sec:error}), but in the interest of transparency, we compute the Fisher uncertainties for all $\xi$ for only scales $20-200\,h^{-1}\mathrm{Mpc}$ in \autoref{sec:error} and \autoref{tab:errscales_xi}. Most changes in error are near $10\%$, although there are a few that exceed this. Fisher forecasts are idealized, however, and so should be trusted up to the $10\%$ level as a rule of thumb.}

The void size function and the halo mass function are uncorrelated with each other, as highlighted by the perpendicular Fisher contours already, confirming that the two probes are strongly complementary. By inspecting the correlation between the void size function and void auto-correlation function, as well as the void size function and the void-halo cross-correlation, we see that large voids tend to be slightly anti-correlated at small scales, while small voids tend to be correlated at small scales. We can also see that large voids tend to be anti-correlated with halos that are close to each other, as expected. Viewing the correlation between the void auto-correlation function and the void-halo cross-correlation function, we see that at scales larger than the void exclusion scale and larger than the void wall rim, respectively, the two functions are somewhat correlated. This is because each follows the matter field at such large scales. We also see a parallel increase in correlation for scales larger than the {\color{black}median} void radius and larger than the void wall rim, due to the fact that voids are more likely to be clustered at these scales. Without the sheer volume of the \gigantes~suite, the correlation among and within these void summary statistics could not be understood in such detail.

\subsection{Optimal constraints combining voids and halos} 

\begin{figure*}[t!]
\centering
\includegraphics[width=0.85\linewidth]{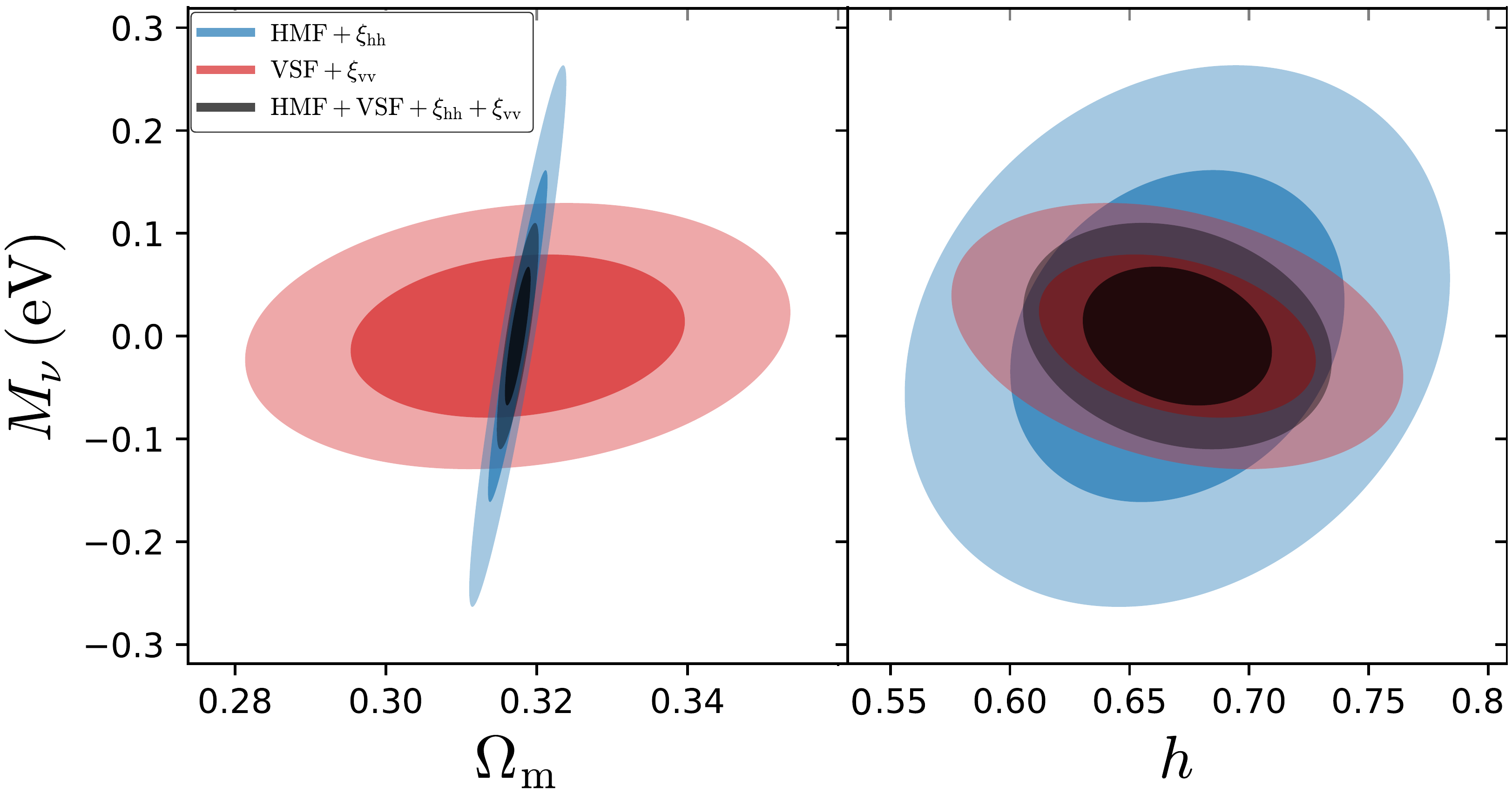}
\caption{Comparison of halo-only with void-only constraints (therefore not including the void-halo cross-correlation).}
\label{fig:compare}
\end{figure*}

The previous section showed that voids provide different, additional information than other probes. Given that voids are obtained \textit{from} the halo or galaxy distribution, we want to compare the total information from all void summary statistics (void size function and $\xi_{\rm vv}$) with the total information from halo summary statistics (halo mass function and $\xi_{\rm hh}$)---without considering the combined statistics $\xi_{\rm vh}$ for this comparison. 
\autoref{fig:compare} shows the comparison {\color{black}for a volume of 1 $h^{-3}\mathrm{Gpc}^3$}: {\color{black}for some parameters, voids in fact perform better than halos, and overall voids are orthogonal in their constraints to halos}. Voids are particularly effective in constraining massive neutrinos and $h$. Constraints on neutrinos are expected to be powerful from voids \citep{Massara_2015, Kreisch_2019}, as diffuse components particularly impact voids. See \autoref{sec:full_constraints} for further parameter comparisons. 

In all cases, as highlighted above, combined constraints are powerful. We recall that for this comparison we have not considered the void-halo cross-correlation, in order to compare halo-only with void-only summary statistics. If adding the void-halo cross-correlation, constraining power is further increased (see \autoref{sec:full_constraints}).
We leave for future work the exploration of the constraining power on dark energy, expected to be large from voids \citep{Bos_2012, Pisani_2015,Verza_2019,Pisani_2019}.

\subsection{Void shape adds information} 
Void definition has been an intense area of discussion in the past \citep{Weygaert_2009,Colberg_2008,Cautun_2018}. Recently, to exploit cosmological information, most works have chosen the approach of testing a particular definition and pushing its use in measuring cosmological signals. This approach has been particularly successful, as the point is not finding the best definition, but merely understanding the sensitivity of a particular definition to the signal to measure. In the realm of maximising the signal {\color{black}and being able to model the signal, which can be more challenging for some void definitions than others}, fully understanding the impact of definition choices is crucial. 

There are various classes of void finders, but a striking differentiating feature is whether the algorithm assumes a spherical shape for voids. While it can be considered intuitive that a void finder with no shape assumption helps when trying to infer cosmological information from the void shape (that is for applications relying on the void-halo cross-correlation function, such as \AP test and RSD measurements from voids, or void ellipticity), it is not so trivial to understand what is the relevance of considering void shape for applications such as the void size function. Our large set of simulations allows us to address this question.

\begin{figure*}[t!]
\centering
\includegraphics[width=0.85\linewidth]{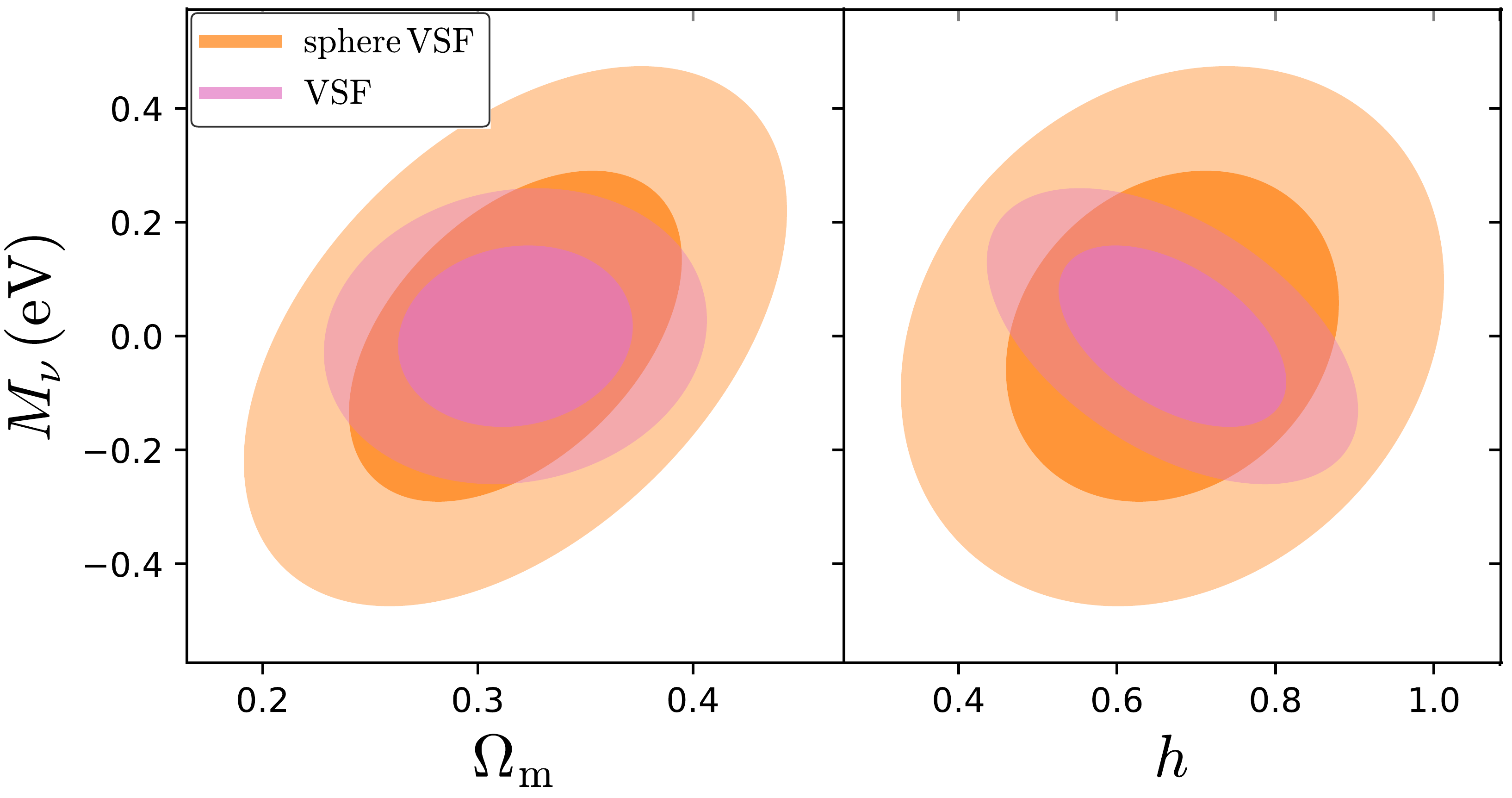}
\caption{Comparison of constraints from a void finder with no prior on the shape of voids, and a more simplistic spherical-assumption based void finder. Voids are found from the same halo field.}
\label{fig:comparesphericalvoids}
\end{figure*}

In this Section we compare the information content captured when the void shape is measured in detail with the case in which a spherical assumption is made by the void finder. In other words we compare constraints obtained when selecting voids with \texttt{\texttt{VIDE}}, a void finder with no prior on void shape, and a more simplistic spherical-assumption based void finder \citep[see e.g.][]{Bayer_2021,Villaescusa-Navarro_2020}. \autoref{fig:comparesphericalvoids} shows results of the comparison {\color{black}for a volume of 1 $h^{-3}\mathrm{Gpc}^3$}. For most of the cosmological parameters considered in this paper the void size function measured by \texttt{VIDE} provides more stringent constraints than the void size function measured by the spherical void finder (see \autoref{sec:full_constraints} for the full set of contours). We notice, however, that in some cases the orientation of constraints is different. This consideration is relevant: in future work aiming to maximise the constraints extracted from the cosmic web, it could be possible to combine constraints from different void finders for the void size function. One has to keep in mind, however, that the correlations between constraints from two different void finders may be non-trivial to be accounted for, but \gigantes~contains enough catalogues to estimate this covariance accurately.

We also note that this comparison is more realistic given that it is obtained with voids found in the halo distribution, and not in the dark matter field. These results showcase for the first time that even for non-shape based applications, such as the void size function, shape plays a strong role in determining the quality of constraints.

\section{Likelihood-free inference of cosmological parameters}
\label{sec:LFI}

In this section we show a machine learning application to the \gigantes~dataset. Our goal is to perform likelihood-free inference from one of the most important summary statistics associated to cosmic voids: the void size function. In order to carry out this task, we need many examples from different cosmological models in order to be able to extract unique patterns that allow us to find a connection between the void size function and the value of the cosmological parameters.

We use the void catalogs from the high-resolution \quijote~latin-hypercube at $z=0$ {\color{black}over a volume of 1 $h^{-3}\mathrm{Gpc}^3$}. We use 18 bins equally spaced from $R=10.5~h^{-1}{\rm Mpc}$ to $R=61.5~h^{-1}{\rm Mpc}$. Our goal is to predict the mean ($\theta$) and standard deviation of the posterior ($\delta \theta$) from the void size function
\begin{equation}
\theta\pm\delta\theta=f({\rm VSF})
\end{equation}
where we are going to use neural networks to approximate the function $f$. We use moment networks \citep{moment_networks} to achieve this.

\begin{figure}[t!]
\includegraphics[width=0.493\linewidth]{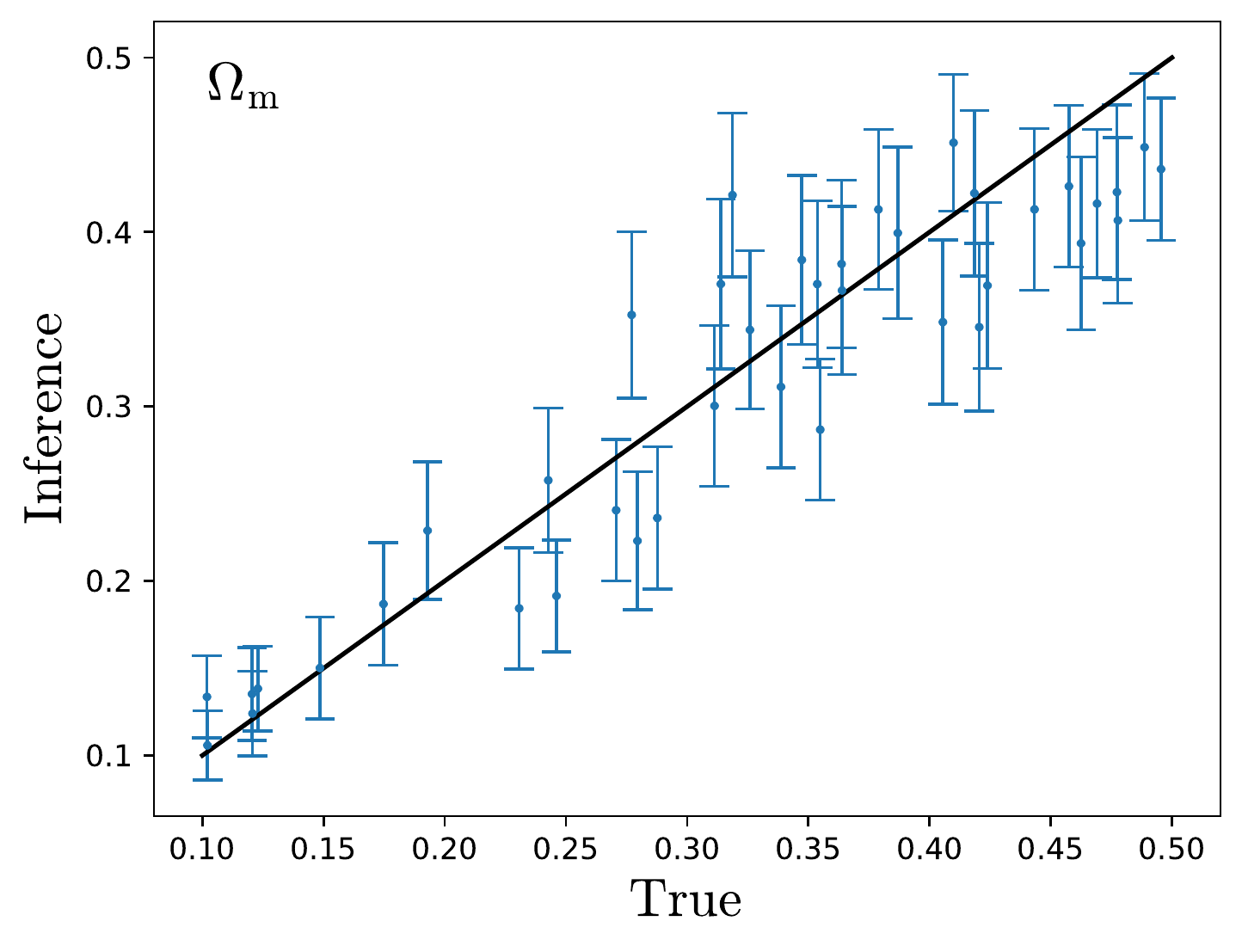}
\includegraphics[width=0.493\linewidth]{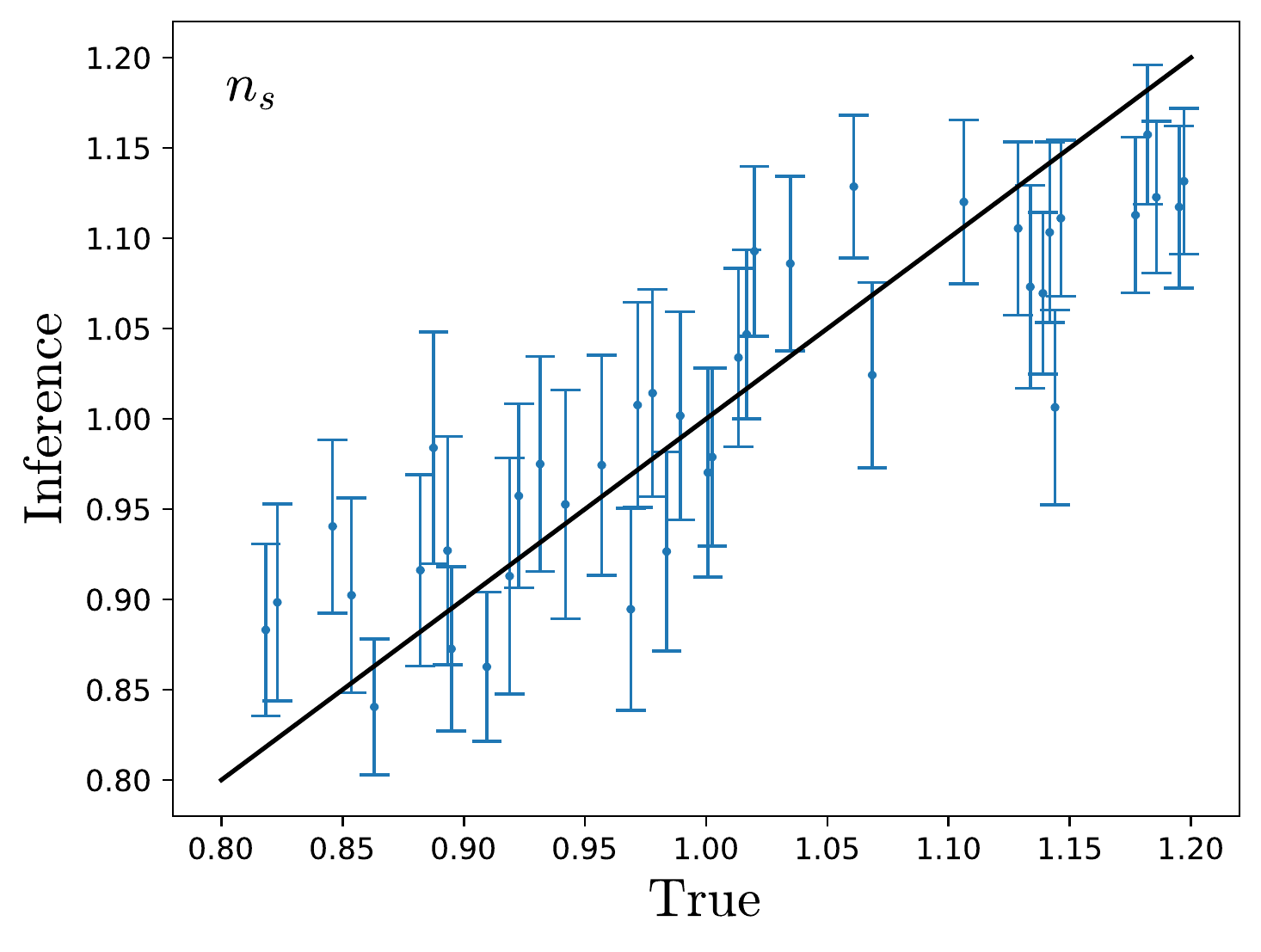}
\caption{We use likelihood-free inference {\color{black}with moment networks \citep{moment_networks}} to estimate the mean and standard deviation of the posterior from measurements of the void size function. The above panels show the results for $\Omega_{\rm m}$ (left) and $n_s$ (right). Each point represents the posterior mean from a measurement of the void size function from one realization of the HR latin hypercube at $z=0$. The error bars display the standard deviation of the posterior. For $\Omega_{\rm m}$ (left) and $n_s$ we are able to recover the true value in most of cases; it is expected that some points lie outside the posterior standard deviation interval. We are unable to recover the value of $\Omega_{\rm b}$, $h$, and $\sigma_8$ from these measurements, as we do not have enough statistical power for these parameters {\color{black} since we only use the VSF from 1$h^{-3}\mathrm{Gpc}^3$}.}
\label{fig:LFI}
\end{figure}

We show the results in \autoref{fig:LFI} for 40 different void size functions of the test set.  We note that we only show results for $\Omega_{\rm m}$ and $n_s$, as those are the only two parameter for which we have enough statistical power to find a relation between the void size function and the value of the parameters. For the case of $\Omega_{\rm b}$, $h$, and $\sigma_8$ our model just predicts the mean with a large errorbar that is not meaningful due to the edges of the prior. 

As it can be seen, our method is able to recover the true value within the standard deviation in most of the cases; it is expected that the true value lies outside of the posterior standard deviation for some cases.

This simple application illustrates the power of the so-called simulation based inference, where accurate and reliable simulations can be used to perform inference over statistics whose likelihood is unknown. We emphasize that this procedure is not limited to summary statistics, but can be carried out at the field level.

\section{Conclusion}
\label{sec:conclusion}

We have introduced the state of the art \gigantes~void catalogs suite obtained from the \quijote~simulations. The catalogs, built with the popular void finder \texttt{VIDE}, provide a comprehensive set of more than 1 billion voids in {\color{black}a large range of $\Lambda\mathrm{CDM}$} cosmologies. 

To present the \gigantes~dataset we show the void size function, average profiles, and the void auto-correlation function. We also compute the correlation among and between different void summary statistics, as well as between traditional quantities such as the halo mass function and the halo auto-correlation function.  

To illustrate the immense power provided by the \gigantes~catalogs, we show for the first time in a realistic set up of voids found in the halo field that voids provide complementary and independent information with respect to halos. Voids allow us to extract higher-order information from the cosmic web that would otherwise remain inaccessible. Our results show that added information is particularly effective with respect to constraining neutrinos and $h$, but helps for most parameters of interest. 

We show that, aside from the void-halo cross-correlation function and the void size function, a strong contribution to constraints comes from the void auto-correlation function. We also compare the power of halos with respect to voids when combining different summary statistics. Finally, by comparing constraints from \texttt{VIDE} voids with voids from a spherical void finder, we show that void shape knowledge provides an important contribution to the overall information content from voids. 

While providing the largest and most realistic set of voids ever released, this paper answers many open questions relevant for void cosmology: Do voids provide additional information? How is that information correlated? Which parameters are optimally constrained by voids? Is void shape important for cosmological constraints?
We have showed that voids allow us to go beyond the two-point correlation function and provide additional constraining power. {\color{black}\gigantes~opens up the exploration of the cosmological constraining power of void statistics beyond the ones considered in this paper.} Theory developments and instrument systematics analyses for void summary statistics will allow us to fully exploit the power of the void size function, the void-halo, the void-void correlation functions{\color{black}, and other void statistics} to constrain cosmological parameters with voids from the next generation of surveys.

\appendix

\section{Simulations and void finder} 
\label{sec:quijote_vide}
\subsection{The \quijote~simulations} 
\label{subsec:quijote}

The \quijote~simulations \citep{Villaescusa-Navarro_2020} are a suite of $44,100$ full N-body simulations exploring more than $7,000$ cosmological models. These simulations contain trillions of particles over a combined volume larger than the volume of the entire observable Universe, and were designed for two main tasks: 1) to quantify the information content on cosmological observables, and 2) to provide enough data to train machine learning algorithms. Given their immense combined volume, they represent a powerful tool to explore the properties of extreme objects, such as galaxy clusters and cosmic voids. The simulations have been used in a range of papers \citep[e.g.][]{Samushia_2021, Bernal_2021, Gualdi_2021, Kuruvilla_2021, Bayer_2021, Obuljen_2019, Banerjee_2019, Changhoon_2019, Giusarma_2019, Uhlemann_2020, Friedrich_2020, Ramanah_2020, Massara_2020, Dai_2020, Philcox_2020, Philcox_Massara_Aviles, Philcox_Massara_Spergel, Philcox_Spergel_Paco, Allys_2020, Miles_2020, Aviles_2020, Parimbelli_2020, Banerjee_2020, Banerjee_2021, Lazeyras_2020, Gualdi_2020, Gualdi_2021, CarPool, Giri_2020, Bella_2020, Renan_2020, Changhoon_2020, Chen_2020, bayer2}. 
On average, each realization of the \quijote~simulations contains $\sim 500,000$ dark matter halos over a volume of 1 $(h^{-1}\mathrm{Gpc})^3$ at $z=0$, corresponding to a halo number density of $4 \times 10^{-4}\,h^{3}\,\mathrm{Mpc}^{-3}$ and mean halo separation of $\sim 13\,h^{-1}\,\mathrm{Mpc}$. The equivalent values at $z=1$ are $2 \times 10^{-4}\,h^{3}\,\mathrm{Mpc}^{-3}$ and $\sim 17\,h^{-1}\,\mathrm{Mpc}$, respectively. Since access to the void hierarchy may depend on tracer number density, it is relevant to notice that the tracer number density of the \quijote~halo catalogs is of a similar order of magnitude than what we can expect from surveys such as PFS \citep{pfs2016}, and denser than DESI \citep{DESI}. With the \quijote~halo catalogs, however, we have a tracer number density below what will be achieved by Euclid \citep{euclid_presentaz}, and the Roman Space telescope \citep{Spergel_2015}, indicating that these upcoming surveys will be able to fully capture the power from the void hierarchy, and can expect even tighter constraints on cosmological parameters than what we illustrate in this work.

\subsection{The void finder \texttt{VIDE}} 
\label{subsec:vide}

\texttt{VIDE} is the most popular Voronoi-watershed based void finder used in Cosmology. Based on \texttt{ZOBOV} \citep{vide:Neyrinck-2008}, \texttt{VIDE} provides a convenient, publicly-released toolkit to find voids in simulations (dark matter/hydro-dynamical) and data (spectroscopic or photometric data). 

The void finder first performs a Voronoi tessellation of the tracer particle field (tracers can be dark matter particles, halos, galaxies), providing a physical splitting of the 3D particle field into cells. Each cell is assigned a density equal to the inverse of the volume of the cells. Starting from local minima (largest cells) of the tessellation, the algorithm creates basins by merging cells with a monotonic increase in density. Through the use of the watershed trasform, \texttt{VIDE} provides the final set voids. 

Void centers are defined as the volume-weighted barycenters of the Voronoi cells composing the void. Void centers are therefore sensitive to the whole structure of the object. Considering a sphere of equivalent volume to the volume $V_i$ of all the cells, one can define an effective radius for voids, $R$: 
\begin{equation}
	R = \left(\frac{3}{4\pi}\sum\nolimits_iV_i \right)^{1/3}\;.
\end{equation}
Even though it provides the effective radius (that we simply refer to as radius in the rest of the paper), \texttt{VIDE} makes no assumption on the void shape, conversely to other void finders assuming sphericity.  

The absence of a shape prior makes \texttt{VIDE} an excellent option for cosmological analyses for which a precise determination of void shape is a strength (e.g. \AP test and redshift-space distortion analyses). Finally, \texttt{VIDE} is built to handle a {\color{black}survey} mask, and can easily be applied to study voids from both data and simulations, as its wide range of applications shows. Among other results, it has been used to provide the most stringent to date constraints from cosmic voids \citep{Hamaus_2020}. Recently, a Python 3 version of the code has been released\footnote{\url{https://bitbucket.org/cosmicvoids/vide_public/wiki/Home}}. 

Since most applications of void finding in data for cosmology rely on Voronoi-based void finders, the work performed in this paper for this category of void finders is pertinent to current cosmological observational analyses. 
The large set of void catalogs presented in this work will be publicly released upon acceptance of the paper.

\section{Fisher Error Stability}
\label{sec:error}

Tables \ref{tab:error_combo}, \ref{tab:error_VSF}, \ref{tab:error_xiv}, \ref{tab:error_xivh}, \ref{tab:error_HMF}, and \ref{tab:error_xihh} show {\color{black}1 standard deviation uncertainties} on cosmological parameters from the denoted summary statistic and given number of simulation realizations {\color{black}marginalized over the parameters listed as well as $\Omega_{\rm b}$}. Percent differences relative to the error with 500 simulations are shown in parentheses. For all void summary statistics, the percent change in error is less than the percent change in the number of realizations, indicating the constraints are stable.

{\color{black}Tables \ref{tab:coverror_combo}, \ref{tab:coverror_VSF}, \ref{tab:coverror_xiv}, \ref{tab:coverror_xivh}, \ref{tab:coverror_HMF}, and \ref{tab:coverror_xihh} also show 1 standard deviation uncertainties, but given the number of simulations used to compute the covariance. Percent differences relative to the error with 15000 simulations are shown in parentheses. Most percent changes are below $1\%$, indicating the covariance is stable. }

{\color{black}Finally, in \autoref{tab:errscales_xi}, we illustrate the impact on the Fisher uncertainties from computing the correlation functions on scales $15-200\,h^{-1}\mathrm{Mpc}$ in log space compared to computing the correlation functions on scales $20-200\,h^{-1}\mathrm{Mpc}$ in log space. Percent differences relative to scales $20-200\,h^{-1}\mathrm{Mpc}$ are shown in parentheses. Most changes are near $10\%$, although a handful are larger. An increase in error is expected as less data are used.}

\section{Full Constraints}
\label{sec:full_constraints}

\begin{figure*}[h!]
\includegraphics[width=\linewidth]{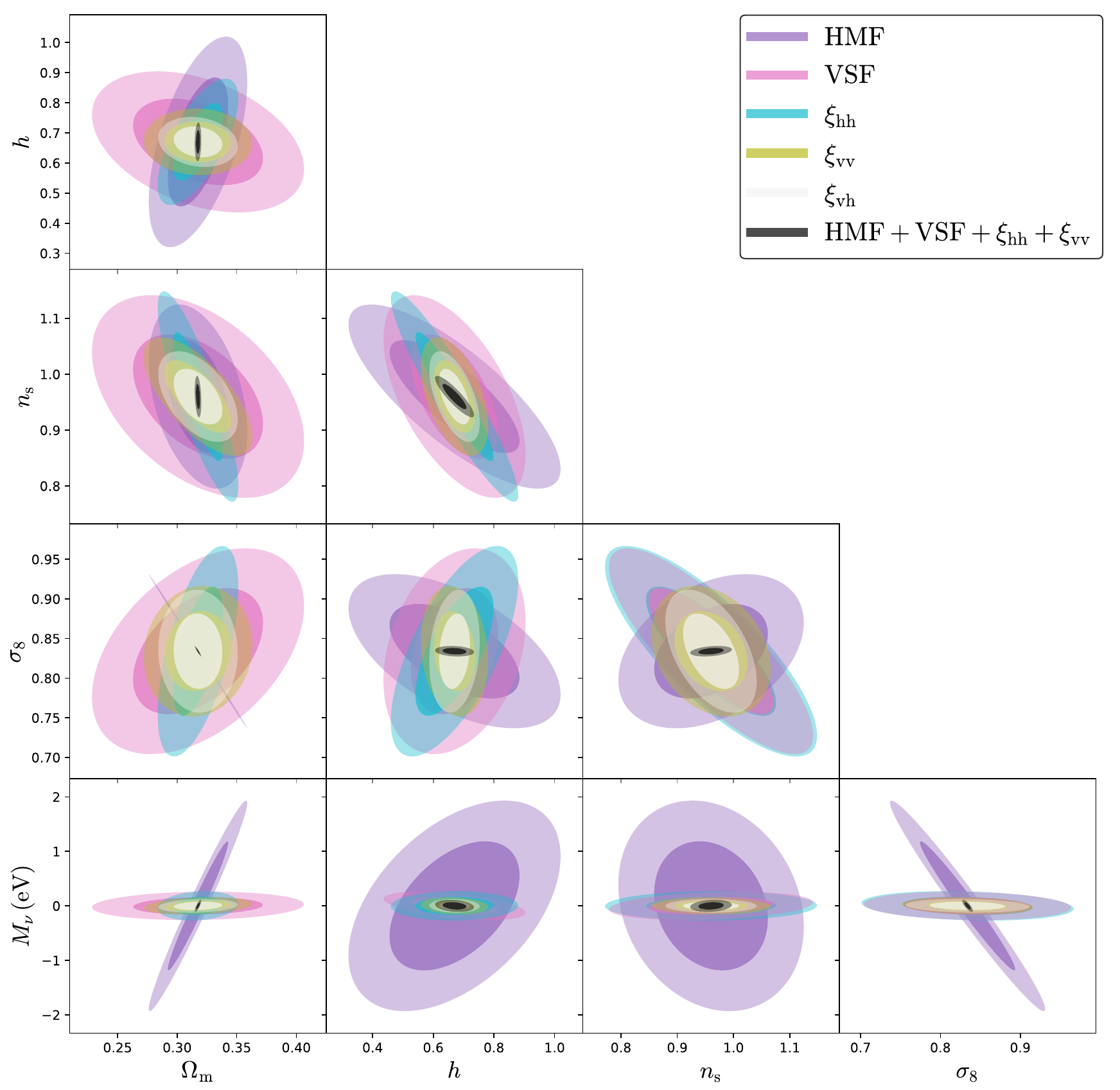}
\caption{Constraints on cosmological parameters from the voids (void size function, void-halo, and void-void correlation functions) and halos (halo mass function, halo auto-correlation function).}
\label{fig:allconstraints}
\end{figure*}

\begin{figure*}[t]
\includegraphics[width=\linewidth]{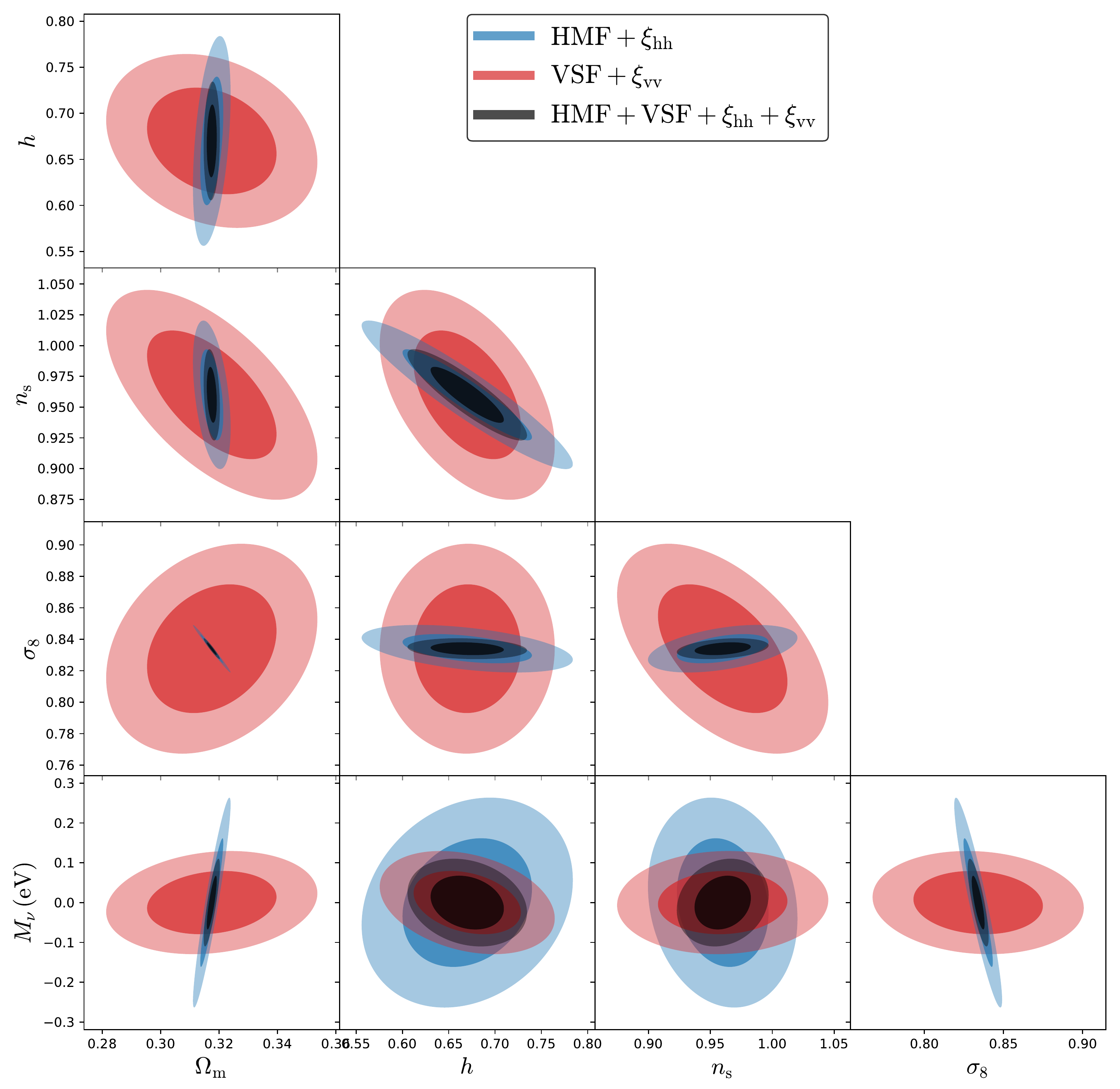}
\caption{Comparison of halo-only with void-only constraints (therefore not including the void-halo cross-correlation) for all parameters.}
\label{fig:allcompare}
\end{figure*}

\begin{figure*}[h!]
\includegraphics[width=\linewidth]{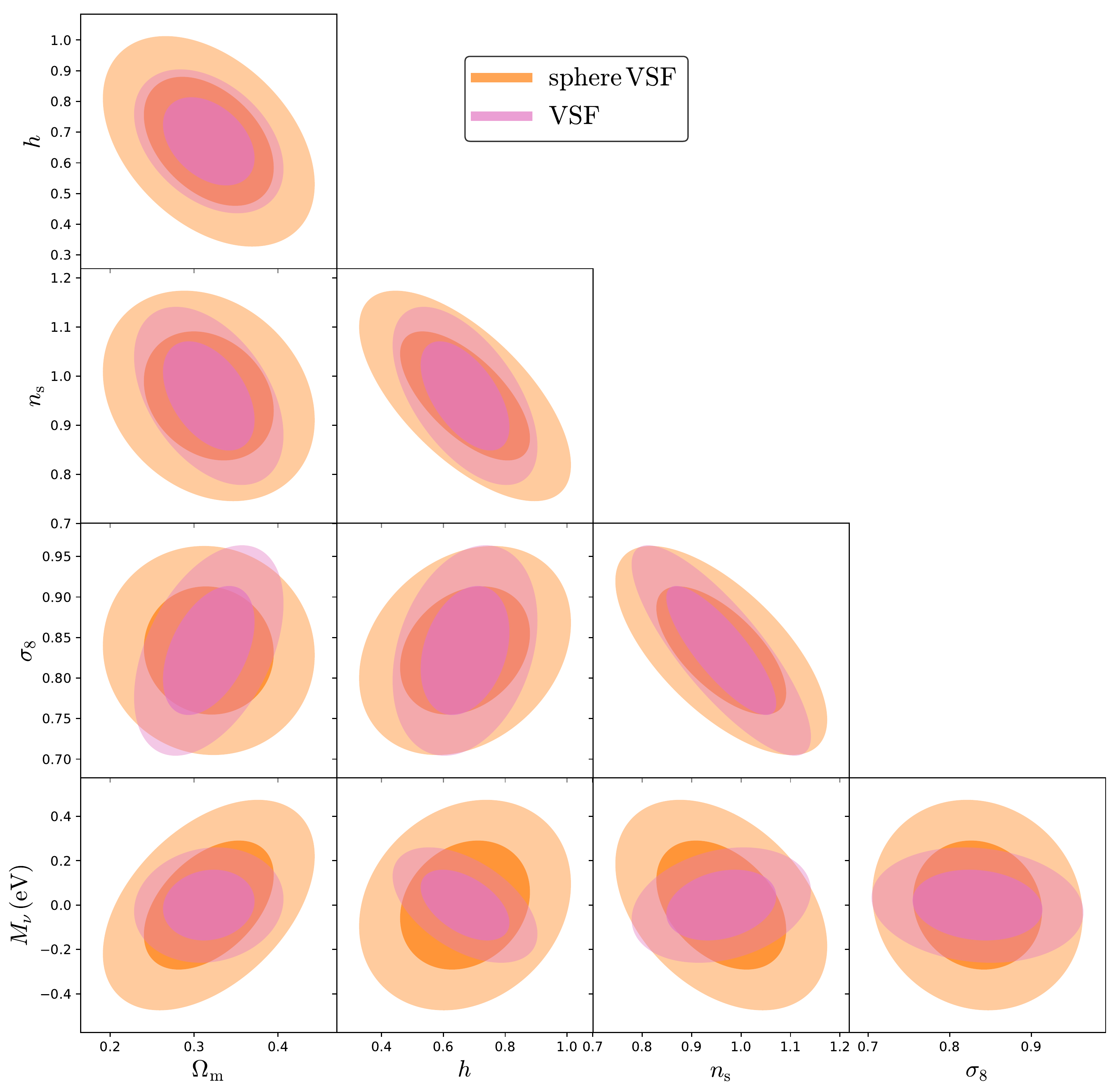}
\caption{Comparison of constraints for all parameters from a void finder with no prior on the shape of voids, and a more simplistic spherical-assumption based void finder. Voids are found from the same halo field.}
\label{fig:allcomparesphericalvoids}
\end{figure*}

We show in \autoref{fig:allconstraints} the constraints from different void summary statistics, as well as for the halo mass function and the halo auto-correlation function, for all the parameters considered in this paper: the sum of neutrino masses, $\sigma_8$, $n_s$, $h$, and $\Omega_\mathrm{m}$. {\color{black}In \autoref{fig:allcompare} and \autoref{fig:allcomparesphericalvoids} we show full cosmological parameter contours for combined halo and combined void summary statistics, and the spherical void finder VSF and VIDE VSF, respectively.}

 \begin{table*}[h]
  \caption{{\color{black}Error Sensitivity for HMF + VSF + $\xi_{\rm vv}$ + $\xi_{\rm hh}$} \label{tab:error_combo}}
  \begin{ruledtabular}
  \begin{tabular}{cccccc}
  Number of Simulations & $\Omega_{\rm m}$ &  $h$ & $n_{\rm s}$ & $\sigma_8$ & $\sum m_{\nu}$\\
  \hline \\ [-2ex]
  500 	&  0.0022		&  0.052 &  0.030 &  0.0054 & 0.089  \\
  450 (10\%) 	& 0.0022  	(0\%) & 0.048   (8\%) & 0.029   (3\%) & 0.0052 (4\%) &  0.086 (3\%) \\
  400 (20\%)	& 0.0022	(0\%) & 0.048  (8\%) & 0.028  (7\%) &  0.0053 (2\%) &  0.083  (7\%)

  \end{tabular}
\end{ruledtabular}
\end{table*}

 \begin{table*}[h!]
  \caption{Error Sensitivity for VSF \label{tab:error_VSF}}
  \begin{ruledtabular}
  \begin{tabular}{cccccc}
  Number of Simulations & $\Omega_{\rm m}$ & $h$ & $n_{\rm s}$ & $\sigma_8$ & $\sum m_{\nu}$\\
  \hline \\ [-2ex]
  500 	& 0.072	&  0.19 & 0.15 & 0.11 & 0.21 \\
  450 (10\%) 	& 0.069	(4\%) & 0.18 (5\%) & 0.14 (7\%) & 0.099 (10\%) & 0.21 (0\%) \\
  400 (20\%)	& 0.062	(14\%) &  0.18 (5\%) & 0.13 (13\%) & 0.094 (15\%) &  0.18 (14\%)

  \end{tabular}
\end{ruledtabular}
\end{table*}

 \begin{table*}[h!]
  \caption{Error Sensitivity for $\xi_{\rm vv}$ \label{tab:error_xiv}}
  \begin{ruledtabular}
  \begin{tabular}{cccccc}
  Number of Simulations & $\Omega_{\rm m}$ &  $h$ & $n_{\rm s}$ & $\sigma_8$ & $\sum m_{\nu}$\\
  \hline \\ [-2ex]
  500 	& 0.037		& 0.089 & 0.086 & 0.067 & 0.13  \\
  450 (10\%) 	& 0.036	(3\%) & 0.082 (5\%) & 0.079 (8\%) & 0.064 (4\%) & 0.12 (8\%) \\
  400 (20\%)	& 0.032	(14\%) &   0.077 (13\%) & 0.074 (14\%) & 0.065 (3\%) &  0.12 (8\%)

  \end{tabular}
\end{ruledtabular}
\end{table*}

 \begin{table*}[h!]
  \caption{Error Sensitivity for $\xi_{\rm vh}$ \label{tab:error_xivh}}
  \begin{ruledtabular}
  \begin{tabular}{cccccc}
  Number of Simulations & $\Omega_{\rm m}$ &  $h$ & $n_{\rm s}$ & $\sigma_8$ & $\sum m_{\nu}$\\
  \hline \\ [-2ex]
  500 	& 0.027		& 0.067 & 0.066 & 0.063 &  0.10 \\
  450 (10\%) 	& 0.026	(4\%) &  0.061(9\%) & 0.063 (5\%) & 0.062 (2\%) & 0.10 (0\%) \\
  400 (20\%)	& 0.024	(11\%) &    0.061 (9\%) & 0.062 (6\%) & 0.062 (2\%) & 0.090  (10\%)

  \end{tabular}
\end{ruledtabular}
\end{table*}

\begin{table*}[h!]
  \caption{ {\color{black}Error Sensitivity for HMF \label{tab:error_HMF}}}
  \begin{ruledtabular}
  \begin{tabular}{cccccc}
  Number of Simulations & $\Omega_{\rm m}$ &  $h$ & $n_{\rm s}$ & $\sigma_8$ & $\sum m_{\nu}$\\
  \hline \\ [-2ex]
  500 	& 0.033	&  0.28 & 0.13 & 0.078 &  1.56 \\
  450 (10\%) 	& 0.027	(18\%) &   0.25 (11\%) & 0.12 (8\%) & 0.064 (18\%) &  1.23 (21\%) \\
  400 (20\%)	& 0.028	(15\%) &    0.26 (7\%) & 0.12 (8\%) & 0.065 (17\%) &  1.27 (19\%)

  \end{tabular}
\end{ruledtabular}
\end{table*}

 \begin{table*}[h!]
  \caption{Error Sensitivity for $\xi_{\rm hh}$ \label{tab:error_xihh}}
  \begin{ruledtabular}
  \begin{tabular}{cccccc}
  Number of Simulations & $\Omega_{\rm m}$ &  $h$ & $n_{\rm s}$ & $\sigma_8$ & $\sum m_{\nu}$\\
  \hline \\ [-2ex]
  500 	& 0.027	&  0.17 & 0.15 & 0.11 &  0.22 \\
  450 (10\%) 	& 0.028	(4\%) &   0.18 (6\%) & 0.16 (7\%) & 0.11 (0\%) & 0.22 (0\%) \\
  400 (20\%)	& 0.029	(7\%) &     0.19 (12\%) & 0.17 (13\%) & 0.11 (0\%) &  0.22 (0\%)

  \end{tabular}
\end{ruledtabular}
\end{table*}

 \begin{table*}[h]
  \caption{{\color{black}Covariance Error Sensitivity for HMF + VSF + $\xi_{\rm vv}$ + $\xi_{\rm hh}$} \label{tab:coverror_combo}}
  \begin{ruledtabular}
  \begin{tabular}{cccccc}
  Number of Simulations & $\Omega_{\rm m}$ &  $h$ & $n_{\rm s}$ & $\sigma_8$ & $\sum m_{\nu}$\\
  \hline \\ [-2ex]
  15000 	&  0.0022	&  0.052 & 0.030  & 0.0054  &  0.089 \\
  13500 (10\%) 	&   0.0022	(0\%) &  0.052  (0\%) &  0.030  (0\%) & 0.0054  (0\%) &  0.089  (0\%) \\
  12000 (20\%)	& 0.0022	(0\%) &  0.052 (0\%) & 0.030  (0\%) &  0.0054 (0\%) &  0.089  (0\%)

  \end{tabular}
\end{ruledtabular}
\end{table*}

 \begin{table*}[h!]
  \caption{\color{black} Covariance Error Sensitivity for VSF \label{tab:coverror_VSF}}
  \begin{ruledtabular}
  \begin{tabular}{cccccc}
  Number of Simulations & $\Omega_{\rm m}$ & $h$ & $n_{\rm s}$ & $\sigma_8$ & $\sum m_{\nu}$\\
  \hline \\ [-2ex]
  15000 	& 0.072 	&  0.19  & 0.15  &  0.10 &  0.21 \\
  13500 (10\%) 	&  0.072	(0\%) &  0.19 (0\%) &  0.15 (0\%) &  0.10 (0\%) &  0.21 (0\%) \\
  12000 (20\%)	& 0.071 	(1\%) &  0.19  (0\%) &  0.15  (0\%) &  0.10  (0\%) & 0.21    (0\%)

  \end{tabular}
\end{ruledtabular}
\end{table*}

 \begin{table*}[h!]
  \caption{\color{black} Covariance Error Sensitivity for $\xi_{\rm vv}$ \label{tab:coverror_xiv}}
  \begin{ruledtabular}
  \begin{tabular}{cccccc}
  Number of Simulations & $\Omega_{\rm m}$ &  $h$ & $n_{\rm s}$ & $\sigma_8$ & $\sum m_{\nu}$\\
  \hline \\ [-2ex]
  15000 	&  	0.037	& 0.089  & 0.086  &  0.067 &  0.13  \\
  13500 (10\%) 	&  0.037	(0\%) &  0.089 (0\%) &  0.085 (1\%) & 0.066  (1\%) &  0.13  (0\%) \\
  12000 (20\%)	&  0.037	(0\%) &   0.089  (0\%) & 0.085  (1\%) & 0.066  (1\%) &  0.13  (0\%)

  \end{tabular}
\end{ruledtabular}
\end{table*}

 \begin{table*}[h!]
  \caption{\color{black} Covariance Error Sensitivity for $\xi_{\rm vh}$ \label{tab:coverror_xivh}}
  \begin{ruledtabular}
  \begin{tabular}{cccccc}
  Number of Simulations & $\Omega_{\rm m}$ &  $h$ & $n_{\rm s}$ & $\sigma_8$ & $\sum m_{\nu}$\\
  \hline \\ [-2ex]
  15000 	&  	0.027	&  0.067 & 0.065  & 0.063  &  0.10  \\
  13500 (10\%) 	&  0.027	(0\%) &  0.067 (0\%) &  0.065 (0\%) &  0.062 (2\%) & 0.10   (0\%) \\
  12000 (20\%)	&  0.027	(0\%) &   0.067   (0\%) &  0.065 (0\%) & 0.062  (2\%) &  0.10  (0\%)

  \end{tabular}
\end{ruledtabular}
\end{table*}

\begin{table*}[h!]
  \caption{ {\color{black}Covariance Error Sensitivity for HMF \label{tab:coverror_HMF}}}
  \begin{ruledtabular}
  \begin{tabular}{cccccc}
  Number of Simulations & $\Omega_{\rm m}$ &  $h$ & $n_{\rm s}$ & $\sigma_8$ & $\sum m_{\nu}$\\
  \hline \\ [-2ex]
  15000 	& 0.033 	&  0.28  & 0.13  & 0.078  &  1.56  \\
  13500 (10\%) 	&  0.033	(0\%) &   0.28  (0\%) &  0.13 (0\%) &  0.078 (0\%) &   1.56 (0\%) \\
  12000 (20\%)	&  0.033	(0\%) &  0.28   (0\%) & 0.13  (0\%) & 0.078  (0\%) &   1.56 (0\%)

  \end{tabular}
\end{ruledtabular}
\end{table*}

 \begin{table*}[h!]
  \caption{\color{black}Covariance Error Sensitivity for $\xi_{\rm hh}$ \label{tab:coverror_xihh}}
  \begin{ruledtabular}
  \begin{tabular}{cccccc}
  Number of Simulations & $\Omega_{\rm m}$ &  $h$ & $n_{\rm s}$ & $\sigma_8$ & $\sum m_{\nu}$\\
  \hline \\ [-2ex]
  15000 	& 0.027 	&  0.17  &  0.15  & 0.11  &  0.22  \\
  13500 (10\%) 	&  0.027	(0\%) &   0.17  (0\%) & 0.15  (0\%) &  0.11 (0\%) &   0.22 (0\%) \\
  12000 (20\%)	& 0.027 	(0\%) &   0.17    (0\%) & 0.15  (0\%) &  0.11 (0\%) &  0.22  (0\%)

  \end{tabular}
\end{ruledtabular}
\end{table*}

 \begin{table*}[h!]
  \caption{\color{black} Impact of Small Scales on $\xi$ Constraints for $\{\Omega_{\rm m}, h, n_{\rm s}, \sigma_8, M_\nu\}$ \label{tab:errscales_xi}}
  \begin{ruledtabular}
  \begin{tabular}{ccc}
  Summary Statistic(s) & 15-200 $\mathrm{h^{-1}\,\mathrm{Mpc}}$ & 20-200 $\mathrm{h^{-1}\,\mathrm{Mpc}}$ \\
  \hline \\ [-2ex]
  $\xi_{\rm vv}$ &  0.037 (30\%), 0.089 (11\%), 0.086 (11\%), 0.067 (17\%), 0.13 (7\%) & 	0.053, 0.10, 0.097, 0.081, 0.14 \\
  $\xi_{\rm vh}$ & 0.027 (10\%), 0.067 (7\%), 0.065 (14\%), 0.063 (6\%), 0.10 (9\%)   & 0.030, 0.072, 0.076, 0.067, 0.11 \\
  $\xi_{\rm hh}$ & 0.027 (10\%), 0.17 (11\%), 0.15 (12\%), 0.11 (27\%), 0.22 (12\%)  & 0.030, 0.19, 0.17, 0.15, 0.25 \\
  HMF + VSF + $\xi_{\rm vv}$ + $\xi_{\rm hh}$ & 0.0022 (12\%), 0.052 (5\%), 0.030 (6\%), 0.0054 (8\%), 0.089 (8\%) &  0.0025, 0.055, 0.032, 0.0059, 0.097  	
  \end{tabular}
\end{ruledtabular}
\end{table*}

\section*{Acknowledgements}
CDK is supported by the National Science Foundation Graduate Research Fellowship under Grant DGE 1656466. 
AP is supported by NASA grant 15-WFIRST15-0008 to the Nancy Grace Roman Space Telescope Science Investigation Team ``Cosmology with the High Latitude Survey'' and NASA ROSES grant 12-EUCLID12-0004.
NH is supported by the Excellence Cluster ORIGINS, which is funded by the Deutsche Forschungsgemeinschaft (DFG, German Research Foundation) under Germany's Excellence Strategy -- EXC-2094 -- 390783311.

\bibliography{paper}{}
\bibliographystyle{aasjournal}

\end{document}